\documentclass[aps, prx, showpacs, twocolumn, superscriptaddress, notitlepage, longbibliography]{revtex4-2}

\usepackage{float}
\usepackage{graphicx}
\usepackage{subfigure}
\usepackage{amsmath}
\usepackage{amssymb}
\usepackage{dsfont}
\usepackage{wasysym}
\usepackage{esvect}

\usepackage{multirow}

\usepackage[xcolor={usenames, dvipsnames}, markup=default]{changes}

\newcommand{\be}{\begin{equation}}
\newcommand{\ee}{\end{equation}}
\newcommand{\ba}{\begin{align}}
\newcommand{\ea}{\end{align}}
\usepackage{hyperref}
\hypersetup{
colorlinks=true,final=true,
        linkcolor=blue,
        citecolor=blue,
        filecolor=blue,
        urlcolor=blue,
}
\newcommand{\beginsupplement}{%
        \setcounter{table}{0}
        \renewcommand{\thetable}{S\arabic{table}}%
        \setcounter{figure}{0}
        \renewcommand{\thefigure}{S\arabic{figure}}
        \setcounter{equation}{0}
        \renewcommand{\theequation}{S\arabic{equation}}%
     }
\newcommand{\nn}{\nonumber}
\newcommand{\cc}{^{\ast}}							% suffix for the complex conjugate
\newcommand{\hc}{^{\dagger}}							% suffix for the hermitian conjugate
\newcommand{\comm}[2]{\left[ #1, #2 \right]} 				% commutator
\newcommand{\ii}{\mathrm{i}}			             			% the imaginary unit i^2=-1
\newcommand{\diss}[1]{\mathcal{D}[ #1 ]}					% dissipator
\newcommand{\Tr}{\text{Tr}}
\newcommand{\ket}[1]{\ensuremath{\left| #1 \right\rangle}}
\newcommand{\E}{\mathcal{E}}				%electric field
\newcommand{\tunneling}{\Delta}  			%the double dot tunneling energy
\newcommand{\detuning}{\epsilon} 			%the double dot detuning ( energy(Right) - energy (Left) )
\newcommand{\qubitAngularFrequency}{\omega_{\rm q}} 		%the qubit energy in frequency units
\newcommand{\resonatorAngularFrequency}{\omega_{\rm r}} 	%the resonator energy in frequency units
\newcommand{\qubitResonatorCoupling}{g} 	%the qubit-resonator coupling strength in energy units
\newcommand{\normCoupling}{\bar g_{\perp}} 	%coupling strength normalized to same frequency
\newcommand{\nGate}[1]{n_{\textrm{G}, #1}} 			% voltage induced gate charge
\newcommand{\eC}[1]{E_{\textrm{C}, #1}}			% charging energy
\newcommand{\capTotal}[1]{ C_{\Sigma,#1} }		% dot total capacitance
\newcommand{\capEqual}{ C_{\Sigma} }	
\newcommand{\capGate}[1]{C_{\textrm{G},#1}}		% gate capacitance
\newcommand{\capMut}{C_\textrm{m}}
\newcommand{\capOut}{C_\textrm{out}}			% dot total capacitance without the capacitance to the other dot
\newcommand{\capOutDot}[1]{C_{\textrm{out},#1}}			% dot total capacitance without the capacitance to the other dot
\newcommand{\capGround}{C_\textrm{gnd}}		% capacitance to ground
\newcommand{\CL}[1]{C_{\textrm{LSG},#1}}%Capacitance to left gate
\newcommand{\CR}[1]{C_{\textrm{RSG},#1}}%Capacitance to right gate
\newcommand{\dVL}[1]{\Delta V_{\textrm{L},#1}}
\newcommand{\dVR}[1]{\Delta V_{\textrm{R},#1}}
\newcommand{\dVLm}{\Delta V_\textrm{m}^\textrm{L}}
\newcommand{\dVRm}{\Delta V_\textrm{m}^\textrm{R}}
\newcommand{\dVLd}{\Delta V_\epsilon^\textrm{L}}
\newcommand{\dVRd}{\Delta V_\epsilon^\textrm{R}}

\begin{document}

\title{\emph{In-situ} Tuning of the Electric Dipole Strength of a Double Dot Charge Qubit:\\ Charge Noise Protection and Ultra Strong Coupling}

\author{P.~Scarlino}
%\altaffiliation{Current address: Hybrid Quantum Circuits Laboratory, EPFL Lausanne, CH-1015 Lausanne, Switzerland}
%\thanks{Hybrid Quantum Circuits Laboratory, EPFL Lausanne, CH-1015 Lausanne, Switzerland}
\affiliation{Department of Physics, ETH Zurich, CH-8093 Zurich, Switzerland} 
\affiliation{Institute of Physics, Ecole Polytechnique F\'ed\'erale de Lausanne, CH-1015 Lausanne, Switzerland}
%\thanks{These authors contributed equally to this work.}
\author{J.~H.~Ungerer}
\altaffiliation{Current address: Swiss Nanoscience Institute, University of Basel, Klingelbergstrasse 82, CH-4056 Basel, Switzerland}
\affiliation{Department of Physics, ETH Zurich, CH-8093 Zurich, Switzerland}
\author{D.~J.~van~Woerkom}
\altaffiliation{Current address: Microsoft Quantum Lab Delft, Delft, 2600 GA, The Netherlands}
\affiliation{Department of Physics, ETH Zurich, CH-8093 Zurich, Switzerland}
\author{M.~Mancini}
\affiliation{Department of Physics, ETH Zurich, CH-8093 Zurich, Switzerland}
\author{P.~Stano}
\affiliation{RIKEN Center for Emergent Matter Science (CEMS), Wako, Saitama 351-0198, Japan}
\affiliation{Department of Applied Physics, School of Engineering,
University of Tokyo, 7-3-1 Hongo, Bunkyo-ku, Tokyo 113-8656, Japan}
\affiliation{Institute of Physics, Slovak Academy of Sciences, 845 11 Bratislava, Slovakia}
\author{C.~M\"uller}
\affiliation{IBM Quantum, IBM Research - Zurich, CH-8803 R\"uschlikon, Switzerland}
\author{A.~J.~Landig}
\affiliation{Department of Physics, ETH Zurich, CH-8093 Zurich, Switzerland}
\author{J.~V.~Koski}
\altaffiliation{Current address: Microsoft Quantum Lab Delft, Delft, 2600 GA, The Netherlands}
\affiliation{Department of Physics, ETH Zurich, CH-8093 Zurich, Switzerland}
\author{C.~Reichl}
\affiliation{Department of Physics, ETH Zurich, CH-8093 Zurich, Switzerland}
\author{W.~Wegscheider}
\affiliation{Department of Physics, ETH Zurich, CH-8093 Zurich, Switzerland}
\author{T.~Ihn}
\affiliation{Department of Physics, ETH Zurich, CH-8093 Zurich, Switzerland}
\author{K.~Ensslin}
\affiliation{Department of Physics, ETH Zurich, CH-8093 Zurich, Switzerland}
\author{A.~Wallraff}
\affiliation{Department of Physics, ETH Zurich, CH-8093 Zurich, Switzerland}
\affiliation{Quantum Center, ETH Zurich, 8093 Zurich, Switzerland}

\date{\today}

\begin{abstract}
	Semiconductor quantum dots, where electrons or holes are isolated via electrostatic potentials generated by surface gates, are promising building blocks for semiconductor based quantum technology.
	Here, we investigate double quantum dot (DQD) charge qubits in GaAs, capacitively coupled to high-impedance SQUID array and Josephson junction array resonators. 
	We tune the strength of the electric dipole interaction between the qubit and the resonator \emph{in-situ} using surface gates. 
	We characterize the qubit-resonator coupling strength, qubit decoherence and detuning noise affecting the charge qubit for different electrostatic DQD configurations. %extracted from stability diagrams, 
	We find that all quantities can be tuned systematically over more than one order of magnitude, 
	resulting in reproducible decoherence rates $\Gamma_2/2\pi<~$5~MHz in the limit of high interdot capacitance.
	Conversely, by reducing the interdot capacitance, we can increase the DQD electric dipole strength, and therefore its coupling to the resonator. 
	By employing a Josephson junction array resonator with an impedance of $\sim4$ k$\Omega$ and a resonance frequency of $\resonatorAngularFrequency/2\pi \sim 5.6$ GHz, we observe a coupling strength of $\qubitResonatorCoupling/2\pi \sim 630$ MHz,
	demonstrating the possibility to achieve the \emph{ultrastrong coupling regime} (USC) for electrons hosted in a semiconductor DQD.
These results are essential for further increasing the coherence of quantum dot based qubits and investigating USC physics in semiconducting QDs.
\end{abstract}
%\pacs{}
\maketitle

The semiconductor material platform~\cite{Hanson2007,zwanenburg2013} promises scalable realizations of quantum bits (qubits) with long coherence time, fast operation, and a wide range of tunability~\cite{Vandersypen2017}. 
Electrons and holes are confined on small islands, called quantum dots (QDs), defined by electrostatic gates fabricated on top of the semiconducting host material~\cite{Wiel2002, Hanson2007, scappucci2020}.
QD devices can be studied directly in transport or remotely by a nearby charge detector, such as a quantum point contact or another quantum dot~\cite{Hanson2007}. 
Recently, semiconducting QDs have also been successfully embedded in a circuit quantum electrodynamics (cQED) architecture,
enabling the study of double~\cite{Frey2012,Petersson2012a} and triple quantum dots~\cite{Landig2018} via their electric dipolar interaction with a microwave resonator. 
Strong coupling between the resonator microwave photons and charge~\cite{Mi2017,Stockklauser2017,Bruhat2018} and spin~\cite{Mi2018,Landig2018,Samkharadze2017} degrees of freedom in the quantum dots has been achieved.
Although the spin degree of freedom is at the focus of attention for quantum information applications,
charge noise in the host substrate remains a major limitation~\cite{dial2013,Yoneda2017b}.
Even operation of the quantum devices at sweet spots~\cite{Petersson2010,Vion2002,Maune2012,Medford2013,Medford2013b,Thorgrimsson2017}--configurations in the parameter space where critical system properties are minimally affected by noise in the control parameters--can only mitigate its effects to a limited extent.
Therefore, understanding and improving the coherence and control of the charge degree of freedom in semiconductor systems is of fundamental interest also for future spin qubit applications. 
In fact, all recent successful cQED implementations for spins of electrons confined in QDs \cite{Mi2018,Landig2018,Samkharadze2017} 
rely on coupling the spin to the electric field of microwave photons via a controlled hybridization of the spin and orbital degrees of freedom, 
in effect allowing the spin qubit to acquire an electric dipole moment.
%hat interacts with the cavity electric field.
The strength of this dipole coupling can be tuned by controlling the spin-orbit degree of hybridization. This allows to identify a compromise between a charge qubit with a short coherence but large coupling to cavity photons 
and the more protected pure spin qubit with small or negligible coupling to cavity photons \cite{Benito2020}.

In this work, we describe a strategy to systematically tune the double quantum dot (DQD) electric dipole strength which controls the coupling rate between the DQD charge system and a superconducting microwave resonator. 
The approach is based on altering the magnitude of the DQD interdot capacitance while maintaining the inter-dot tunneling rate close to the resonator frequency. 
We explore different configurations of the DQD confinement potential created by the surface metallic depletion gates, and demonstrate how to efficiently asses the magnitude of the DQD dipole strength
in a given configuration.

\begin{figure}[!b]
\includegraphics[width=\columnwidth]{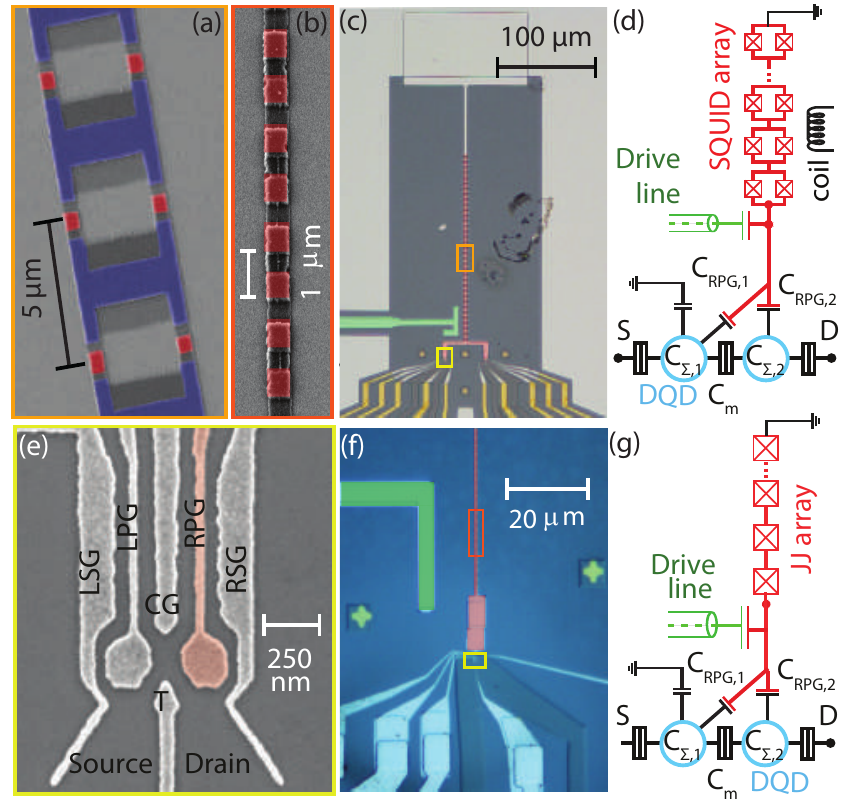}
\caption{\textbf{Simplified circuit diagram and micrographs of the devices.} 
	(a) [(b)] False colored SEM micrograph of a section of the SQUID [Josephson junction] array resonator indicated by the light [dark] orange rectangle in panel (c) [(f)]. The Josephson junctions in the array are highlighted in red.
	(c) False colored optical micrograph of the measured device described in Sec.II, with a SQUID array resonator (red), ground plane (light grey), fine (light grey) and coarse (gold) gates defining the DQD. 
	(d) [(g)] A schematic of the device and control line indicating a simplified circuit diagram of the SQUID [Josephson junction] array resonator (red), drive line (green), the DQD (cyan) and an external coil (black). 
	$C_{\rm{RPG},2}$, $C_{\rm{RPG},1}$, $C_{\rm{\Sigma},2}$, $C_{\rm{\Sigma},1}$ and $C_{\rm{m}}$ are the capacitance between the $\rm{QD}_2$ [$\rm{QD}_1$] and the resonator, total capacitance of $\rm{QD}_2$ [$\rm{QD}_1$] and interdot capacitance, respectively.
(e) Scanning electron micrograph of the areas indicated by yellow rectangles in panels (c) and (f) showing the DQD fine gates (light grey) on the GaAs mesa (dark grey). 
	The plunger gate galvanically connected to the resonator is highlighted in red.
	(f)  False-colored optical micrograph of the measured device described in Sec.III, showing the substrate (dark blue), the superconducting structures including the Al fine gate forming the DQD (light blue), the Josephson junction array (red) and the microwave feedline (green). 
	}
\label{fig1:SampleAndCircuit}
\end{figure}

In this manuscript we present experiments on two distinct devices (reported in Fig.\ref{fig1:SampleAndCircuit}(c) and (f), respectively) with which we explore a range of the DQD electric dipole strength and analyze the DQD decoherence, sensitivity to charge noise, and coupling to the resonator.

In a set of experiments performed with the first device (see Fig.\ref{fig1:SampleAndCircuit}(c))
we systematically decrease the DQD electric dipole strength by exploring regimes in which the interdot mutual capacitance $C_{m}$ becomes the dominant contribution to the DQD capacitance.
This allows us to generate a high degree of resilience against charge noise. 
In this set of experiments we make extensive use of the frequency tunability of the SQUID array resonator (see Fig.\ref{fig1:SampleAndCircuit}(a,d)).
We reproducibly achieve a decoherence rate of only a few MHz for DQD charge qubits in GaAs/AlGaAs operated in the tens of electrons regime \cite{Scarlino2017b}.
This substantially increases the visibility of the vacuum Rabi mode splitting for a DQD-resonator hybrid device, essential for characterizing spectroscopically the coherent electron-photon hybridization.
Furthermore, we show that this reduced sensitivity to charge noise also considerably increases the qubit coherence even at finite DQD detuning.

In a second set of experiments, making use of different device with a Josephson junction array resonator
(see Fig.\ref{fig1:SampleAndCircuit}(b,g)), we explore the same tuning strategy of the DQD confinement potential used in the first set of experiments, but stive for maximizing the DQD electric dipole strength. 
In this way, we increase
the coupling rate of the DQD to the microwave resonator and achieve the Ultra Strong Coupling (USC) regime~\cite{Niemczyk2010, Kockum2018a, FornDiaz2019}. The latter is a configuration
where the vacuum Rabi frequency ($g$) becomes an appreciable fraction of the uncoupled eigenfrequencies of the system ($\omega_{\rm{r}}$), frequently characterized by the ratio $\qubitResonatorCoupling /\resonatorAngularFrequency \geq 0.1$.
In the USC regime, the routinely invoked rotating-wave-approximation is no longer applicable, and the anti-resonant terms become significant, in contrast to standard cavity-QED scenarios~\cite{Kockum2018a, FornDiaz2019}.
Given the smaller electric dipole moment and typically high decoherence rates, reaching the USC regime with a semiconductor DQD system is more demanding than with the superconducting qubits.
Here, we demonstrate that careful design and tuning of the DQD confinement potential
and using a junction array resonator with a characteristic impedance of $\sim 4\,\rm{k}\Omega$ allows us to reach a coupling strength of $\qubitResonatorCoupling /2\pi \sim 600-650$ MHz at a resonator frequency of $\resonatorAngularFrequency/2\pi \sim 5.6~ \rm{GHz}$. 

The article is structured as follows: 
In Sec.~\ref{sec:DQD} we discuss the double quantum dot charge qubit and derive its sensitivity to applied voltages and charge fluctuations, which is central to the understanding of the experiments  presented in later sections.
In Sec.~\ref{sec:Coherence} we present measurements aimed at maximizing coherence of semiconductor charge qubits. 
In Sec.~\ref{sec:USC} we demonstrate that in a device with an identical quantum dot design, we can reach ultra-strong coupling to a superconducting resonator.
We conclude with Sec.~\ref{sec:Conclusion} where we also give an outlook towards future research enabled by these results. Technical details, derivations, and supporting measurements are discussed in the Appendix.

\section{Double quantum dot charge qubit\label{sec:DQD}}
In this work, we consider a double quantum dot charge qubit \cite{vanderwiel2002} coupled to a microwave resonator. We investigate its coherence properties and coupling strength when systematically varying the electrostatic properties of the dot. 
The qubit is modeled by two parameters, the detuning between the two dots $\detuning$ and the tunneling amplitude $\tunneling$ coupling them, in the Hamiltonian
\begin{align}
	H_{\rm{q}} = \frac{1}{2} \Big( \detuning \sigma_z + \tunneling \sigma_x \Big) \equiv \frac{1}{2} \hbar \qubitAngularFrequency \boldsymbol{\sigma} \cdot ( \cos \varphi, 0, \sin \varphi)\,.
	\label{eq:HDQD}
\end{align}
Here, we introduced the mixing angle through $\tan{\varphi} = \detuning / \tunneling$, the qubit energy $\hbar \qubitAngularFrequency = \sqrt{\detuning^2 + \tunneling^2}$ and the vector of Pauli matrices $\boldsymbol\sigma$.
The Hamiltonian is written in the basis of position states $\ket{l}$ and $\ket{r}$, 
which differ in their charge configuration by a single electron transferred across the double dot.
The finite overlap of these position states results in the tunneling amplitude $\tunneling$, and their energy difference defines the detuning $\detuning = \epsilon_r - \epsilon_l$.

The DQD is defined through electrostatic gates controlled via applied voltages. Its states can be characterised by the number of charges in each dot. 
We define a vector of charges $\mathbf{q} = -e\: (n_{1}, n_{2})^{T}$ and gate voltages $\mathbf{v}$. 
The latter leads to induced gate charges on each dot through $\mathbf{q}_{G} = -e\: (\nGate{1}, \nGate{2})^{T} = -e\: \mathds{C}_{G} \mathbf{v}$, 
with the gate capacitance matrix $\mathds{C}_{G}$ and the electron charge $e$ (for details see Appendix \ref{app:sensitivity}).
For a given charge-voltage configuration, the electrostatic energy of the DQD is then given by \cite{Hanson2007}
\begin{align}
	E(n_{1},n_{2},\mathbf{v}) = \frac12 (\mathbf q -\mathbf q_{G})^{T} \cdot \mathds{C}_{D}^{-1} \cdot (\mathbf q - \mathbf q_{G}) \,.
\label{eq:chargingenergy}
\end{align}
Here, we introduced the DQD capacitance matrix 
\begin{align}
	\mathds C_{D} =\begin{pmatrix}
					\capTotal{1} & -\capMut \\
					-\capMut & \capTotal{2}\\
				\end{pmatrix} \,, 
\label{eq:dotCapacitanceMatrix}
\end{align}
with the total capacitance of the $k$-th dot $\capTotal{k}$ and the mutual inter-dot capacitance $\capMut$.
Importantly, the mutual capacitance $\capMut$ is a parameter which is experimentally tuneable, through modifications of the shape and distance of the two dots.

The detuning $\detuning$ in the Hamiltonian is defined as the energy difference between two states whose charge configuration differs by a single charge either on the left or right dot. 
We can thus write
\begin{align}
	\detuning =& E(n_{1}, n_{2} ,\mathbf{v}) - E(n_{1}-1, n_{2}+1 ,\mathbf{v}) \nn\\
		=& \eC{1} (2n_{1} - 2 \nGate{1} -1) - \eC{2} (2n_{2} - 2 \nGate{2} +1) \nn\\
		&+ 2 \eC{\text{m}} (n_{2} - \nGate{2} - n_{1} + \nGate{1} +1)\,, 
%	\detuning =& E(n_{1} - 1, n_{2} + 1 ,\mathbf{v}) - E(n_{1}, n_{2} ,\mathbf{v}) \nn\\
%		=& - \eC{1} (2n_{1} - 2 \nGate{1} - 1) + \eC{2} (2n_{2} - 2 \nGate{2} + 1) \nn\\
%		&+ 2 \eC{\text{m}} (n_{1} - \nGate{1} - n_{2} + \nGate{2} - 1)\,, 
	\label{eq:detuning}
\end{align}
where we defined the charging energies $\eC{1/2} = e^{2} \capTotal{2/1}/[2 (\capTotal{1} \capTotal{2} - C_{m}^{2})]$ 
and $\eC{\text m} = e^{2} C_{m}/[2 (\capTotal{1} \capTotal{2} - C_{m}^{2})]$.
To elucidate the effect of variations and fluctuations in gate voltages $\delta V_{G}$ on the Hamiltonian parameters, we 
define the induced variation in gate charge as $\delta \mathbf q_{G} = \delta V_{\rm G} (\capGate{1}, \capGate{2} )^{T}$.
From Eq.~\eqref{eq:detuning}, we then find the change in $\detuning$ as
\begin{align}
%	\delta \detuning =& 2 \delta V_{\rm G} \left[ \capGate{2} (\eC{2} - \eC{\text m}) - \capGate{1} (\eC{1} - \eC{\text m}) \right] \nn\\
%		\approx& \frac{e^{2} \: \delta V_{\rm G}}{\capEqual + C_{m}} \left( \capGate{2} - \capGate{1} \right).
	\delta \detuning =& 2 \delta V_{\rm G} \left[ \capGate{1} (\eC{1} - \eC{\text m}) - \capGate{2} (\eC{2} - \eC{\text m}) \right] / e\nn\\
		\approx& \frac{e \: \delta V_{\rm G}}{\capEqual + C_{m}} \left( \capGate{1} - \capGate{2} \right) \,,
	\label{eq:sensitivity}
\end{align}
where in the last step, 
we assumed equal QDs with $C_{\Sigma,1} = C_{\Sigma,2} = C_{\Sigma}$. 
The generalization of Eq.~\eqref{eq:sensitivity} to the case of dissimilar QDs is given in Appendix C. 

We will show that qubit electrical sensitivity, expressed in Eq.~\eqref{eq:sensitivity}, appears as an essential parameter for both qubit-resonator coupling and coherence. 
Let us, therefore, analyze  Eq.~\eqref{eq:sensitivity} in more detail. 
It states that the sensitivity to a given gate voltage is larger, if the two dots are coupled to it differently, $\capGate{1} \neq \capGate{2}$, and is smaller if the dot mutual capacitance $\capMut$ grows. 
The more tightly the two dots are coupled, the less differently they respond to a voltage change on a gate and the smaller is the double dot effective dipole strength. 
This is a central point of this manuscript. 

\begin{figure*}[!ht]
\includegraphics[width=\textwidth]{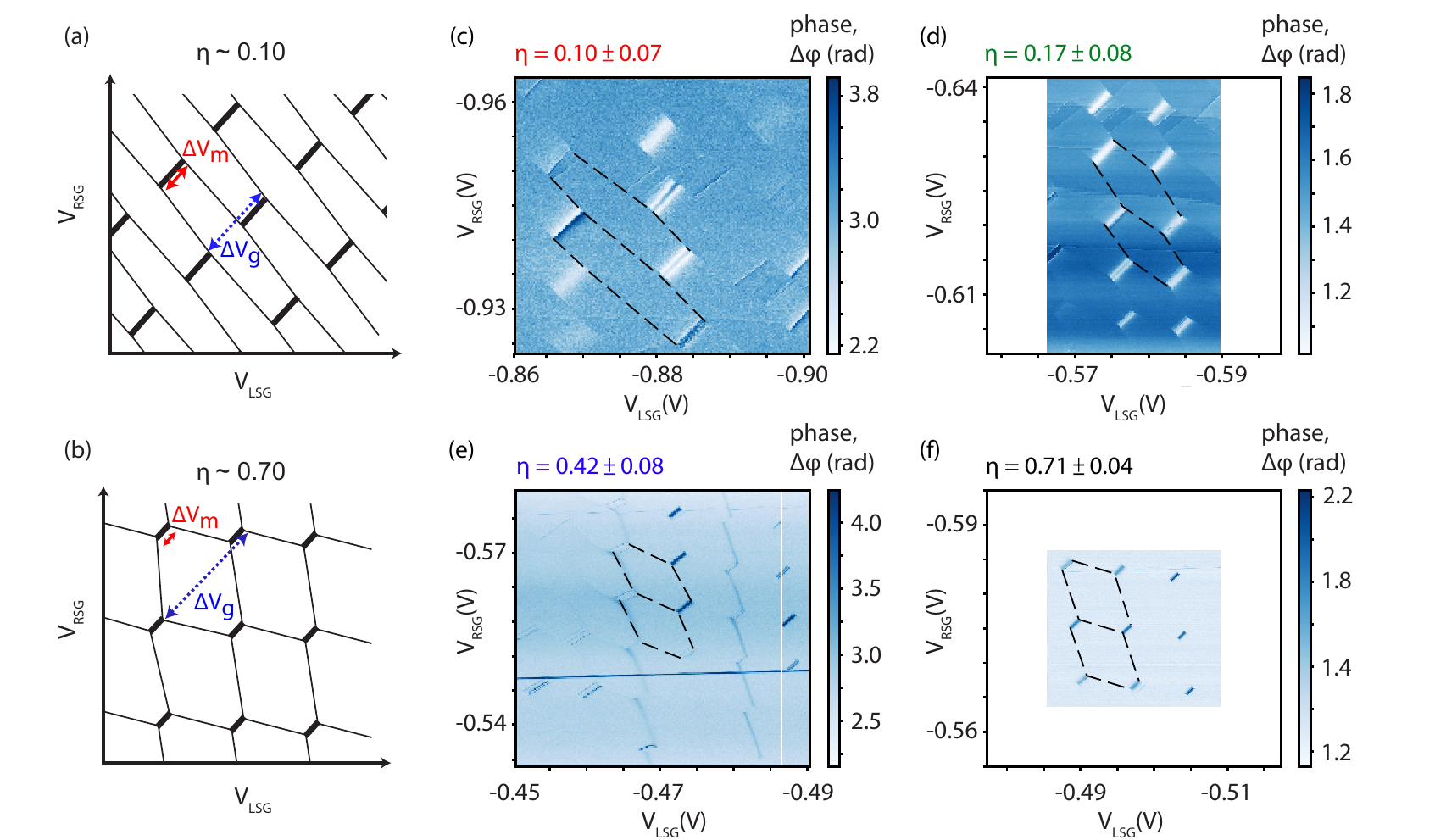}
	\caption{
	%\cm{Caption needs adjusting now - especially how do the indicated voltages translate into the capacitances?}
		(a) A schematic of a DQD charge stability diagram for a configuration with a large mutual capacitance $C_{m}$, resulting in $\eta\sim0.10$.
		The grey areas (black lines) in the charge stability diagram represent interdot (QD$_i$-lead$_i$) charge degeneracy regions. 
		The dipole strength $\eta$ is determined directly from the charge stability diagrams. $\Delta V_{m}$ and $\Delta V_{g}$ are the voltage distance between the two triple points and QD-lead energy degeneracies, respectively. 
		(b) Same as (a), but for smaller $C_{m}$, resulting in $\eta \sim 0.70$.
		(c-f) measured DQD charge stability diagrams obtained for four different DQD configurations in correspondence of four distinct values of $C_m$ [decreasing from panel (c) to (f)]. 
	Each charge stability diagram is measured by monitoring the change in the phase $\Delta\phi$ of the resonator reflectance in response to the DQD gate voltages. 
	The magnitude of the range of the LSG and RSG gate voltages explored in each measurement, $\Delta V_{LSG}$ and $\Delta V_{RSG}$, is kept the same in the four panels for ease of comparison.
	}
	\label{fig2:four_stability_diagramss}
\end{figure*}

However, the reduction is stronger than the factor $1/(\capEqual + \capMut)$ in Eq.~\eqref{eq:sensitivity} would imply on the first look, due to a sum rule that the capacitances need to satisfy. To see that, we write a single dot total capacitance as%~\cm{rethink notation, 
\begin{align}
	\capEqual = \capMut + \capGround + \sum_{\rm g} C_{\rm g} = \capMut + \capOut\,,
	\label{eq:sumRule}
\end{align}
where we define its capacitance to ground as $\capGround$, and to each gate $C_{\rm g}$.
 We also used $\capOut$, the capacitance to the outside world, as the total capacitance to everything else except of the other single dot. With this notation, we write the variation of $\detuning$ due to an applied voltage $\delta V_{G}$ as
\begin{align}
	\delta \detuning = e\: \delta V_G \, \frac{\capGate{1} - \capGate{2}}{\capOut} \frac{\capEqual - \capMut}{\capEqual + \capMut} \,.
	\label{eq:sensitivity2}
\end{align}
%where we again assumed equal dots with $C_{\rm out,1} = C_{\rm out,2} = C_{\rm out,k}$.
Here, we interpret the last term as the renormalization factor for the dipolar energy of the system
\begin{align}
	\eta = \frac{\capEqual - \capMut}{\capEqual + \capMut} =  \frac{1 - \capMut / \capEqual}{1+ \capMut / \capEqual} \,.
	\label{eq:suppression}
\end{align}
If the dots are not equal, an additional contribution appears in Eq.~\eqref{eq:sensitivity2}. However, the definition of the factor $\eta$ given in Eq.~\eqref{eq:suppression} remains the same, see App.~\ref{app:sensitivity} for details.
In the rest of the manuscript we refer to this quantity as \emph{dipole strength} for brevity.  
The quantities defining $\eta$ as given in Eq.~\eqref{eq:suppression} can be directly read off the standard charging diagram of the double dot, as we illustrate in Fig.~\ref{fig2:four_stability_diagramss} and Fig.~\ref{fig:extract eta}.

Note that here we are not considering the concommitent change in tunneling amplitude $\tunneling$ when changing the electrostatic confinement of the dot. 
This is because the lever arm for changing the tunneling amplitude $\tunneling$ in GaAs quantum dots similar to the one considered here is typically at least one order of magnitude smaller than for changes in $\detuning$ \cite{Paladino2014}. 
Furthermore, in the experiments presented here, through independent tuning of the T and CP gate voltages [see Fig.~\ref{fig1:SampleAndCircuit}(e)], we take care to keep $\tunneling$ around $4.5-5.5 \, \rm{GHz}$ in all measurements (see Table \ref{tab1:para_fig3_large}). In this way, 
we can specifically investigate changes in coherence properties and coupling strength when tuning the interdot capacitance $\capMut$ and therefore only the dipole strength $\eta$.

Equations \eqref{eq:sensitivity2} and \eqref{eq:suppression} allow a straightforward derivation of the interaction of the charge qubit and the resonator,
replacing the voltage fluctuations $\delta V_{\rm G}$ by 
\begin{equation}
	\delta V_{\rm G} = \sqrt{\frac{\hbar\resonatorAngularFrequency}{2C_{\rm r}}}(a+a^{\dagger}),
\end{equation}
{i.e.,} the voltage drop of a quantized LC circuit, being here a superconducting resonator of frequency $\resonatorAngularFrequency$. 
The circuit resonant frequency $\resonatorAngularFrequency = 1/\sqrt{L_{\rm r} C_{\rm r}}$ is given by its capacitance $C_r$ and inductance $L_r$
and $a$ is the annihilation operator of its quantized electromagnetic field. 
The strength of the resulting qubit-resonator interaction %(estimated at $\detuning=0$).
$H_{\rm q-r} = (1/2) \qubitResonatorCoupling \sigma_z(a+a^{\dagger})$ can be parameterized using the resonator impedance $Z_{\rm r} = \sqrt{L_{\rm r} / C_{\rm r}}$ as
\begin{align}
	g = \hbar \resonatorAngularFrequency \sqrt{\frac{2 e^2}{\hbar} Z_{\rm r}} \times \eta \: \frac{\capGate{1} - \capGate{2}}{C_{\rm out}} \,,
	\label{eq:coupling}
\end{align}
separating the contributions from the resonator and the DQD charge qubit. 
Since instrumental constraints typically limit the resonator frequency from above, the crucial resonator parameter when aiming at maximizing the coupling strength is its impedance $Z_r$. 
The dot properties and system geometry enter through the second term. 
Section~\ref{sec:USC} demonstrates maximizing the coupling strength $g$ considering both terms.

Equation~\eqref{eq:sensitivity} also encodes the qubit coupling to electrical noise. 
To describe that, we consider uncontrolled fluctuations of voltage $V_{\rm G}$, causing random fluctuations of the qubit energy and thus decoherence. 
The latter is a complex process, depending on the details of the time correlations in these fluctuations. 
After analyzing most typical scenarios~\cite{Ithier2005}, which we list in Appendix \ref{app:scaling}, 
here we restrict ourselves to dephasing due to singular noise with a $1/f$-type spectral function $S(\omega) = A / |\omega|, \omega_{\rm ir} < \omega < \omega_{c}$, linearly coupled to the qubit. 
The low- and high-frequency cutoffs $\omega_{\rm ir}$ and $\omega_{\rm c}$ are typically defined through experimental timescales.
In the quasi-static approximation, where the noise is considered static in each individual run of the experiment, this leads to decay of the qubit off-diagonal density matrix element with a Gaussian form~\cite{Ithier2005} as
\begin{align}
	\ln c_{\rm lin}^{1/f}(\tau)  &=  -\tau^2 \left(\frac{\partial \hbar \qubitAngularFrequency}{\partial \detuning} \right)^2 \left(\frac{\partial \detuning}{\partial V_{\rm G}}\right)^2 
			 A \ln{\left(\frac{\omega_{\rm c}}{\omega_{\rm ir}}\right)} \nn\\ 
	 	& \equiv  -\tau^2 \left(\frac{\partial \hbar \qubitAngularFrequency}{\partial \detuning} \right)^2
				\sigma_{\detuning}^2
				\equiv  -(\Gamma_\varphi \tau)^2.
	\label{eq:decoherence}
\end{align}
Here $\tau$ is the evolution time, and $c(\tau)$ is the decay envelope. 
Writing the expression as a function of a dimensionless parameter $\Gamma_\varphi \tau$, after the second equality sign, defines the pure dephasing rate $\Gamma_\varphi$. 
More importantly, the noise parameter $\sigma_{\detuning} \propto \partial\detuning / \partial V_{\rm G}$, given by Eq.~\eqref{eq:sensitivity}, isolates the effects that are in our focus. 
In Section~\ref{sec:Coherence}, we illustrate how to use these effects to optimize coherence.
Finally, the noise of semiconducting charge qubits is most probably not dominated by fluctuating voltages of the gates, but fluctuating charges of impurities. 
We show in Appendix \ref{app:sensitivity} that there is a formula analogous to Eq.~\eqref{eq:sensitivity2} describing detuning response to a charge impurity fluctuation. 

The dipole strength as defined in Eq.~\eqref{eq:suppression} is experimentally easily accessible and provides useful qualitative predictions. 
Indeed, from Eq.~\eqref{eq:coupling} we see that the coupling to the resonator scales proportionally to $\eta$. Maximizing the coupling therefore calls for maximizing $\eta$, i.e. minimizing the mutual capacitance of the two dots. 
We report on experiments in this regime in Sec.~\ref{sec:USC}. 
In the hypothesis that the coherence of the DQD charge system is limited by electric noise-induced dephasing,
the coherence time $1/\Gamma_{2} \sim 1/\Gamma_{\rm{\phi}}$ is, according to Eq.~\eqref{eq:decoherence}, expected to scale as $1/\eta$, since Eqs.~\eqref{eq:sensitivity2} and \eqref{eq:suppression} give $\partial \detuning / \partial V_{\rm G} \sim \eta$.
A maximally coherent charge qubit, therefore, requires minimizing $\eta$.
The scaling $1/\eta$ is a consequence of the singular noise resulting in a Gaussian decay form. 
Other relevant decay channels, like relaxation and non-singular noise, will lead to a scaling of the coherence time as $\propto 1/\eta^{2}$~\cite{Ithier2005}.
We thus expect that depending on the details of the dominant noise source in the experiments, the qubit coupling quality factor $Q = \qubitResonatorCoupling / \Gamma_{2}$ is either constant as a function of $\eta$ (for singular noise dominating dephasing) or 
can be $\propto 1/\eta$ (for regular dephasing noise or if relaxation dominates). The latter situation would allow one to optimize $Q$ by tuning the mutual dot capacitance.
In the following Sec.~\ref{sec:Coherence}, we investigate which specific scenario is realized in our system.
%We confirm this in Sec.~\ref{sec:Coherence}. 

\section{Increasing charge qubit coherence \label{sec:Coherence}}
In a first set of experiments,
we investigate a GaAs DQD charge qubit strongly coupled to a SQUID array resonator [see Fig.~\ref{fig1:SampleAndCircuit}(a,c)]~\cite{Stockklauser2017,Scarlino2017b} and characterize the qubit coherence properties and its coupling strength $\qubitResonatorCoupling$ to the resonator. 
Aiming to reduce decoherence of the qubit, we \textit{in-situ} explore different electrostatic confinement potentials of the DQD in the few-electron regime ($\sim 10-20$) in the same device by tuning the voltages applied to the electrostatic gates defining the DQD [Fig.~\ref{fig1:SampleAndCircuit}(b)]. 
Each configuration leads to a different strength of the effective dipole interaction between DQD and resonator, characterized by a different dipole strength $\eta$ as defined in Eq.~\eqref{eq:suppression}. 

We use a GaAs/AlGaAs heterostructure with a 2DEG $\sim 90~\rm{nm}$ below the surface. 
Depletion gates are used to define the DQD electrostatic potential. 
The right dot plunger gate is galvanically connected to the resonator [see Figs.~\ref{fig1:SampleAndCircuit}(c-e)]. 
We measure the resonator response in reflection via the drive line [indicated in green in Fig.~\ref{fig1:SampleAndCircuit} (b,c)] in a heterodyne detection scheme by monitoring the amplitude ($|S_{11}|$) 
and phase difference ($\Delta\phi=\mathrm{Arg}[S_{11}]$) of the reflected signal \cite{Wallraff2004}. 
An additional spectroscopy tone can be applied through the same line. The second DQD in the device [Fig.~\ref{fig1:SampleAndCircuit}(c)] is tuned deeply into Coulomb blockade and does not participate in the reported experiment.
\begin{figure*}[!t]
\includegraphics[width=\textwidth]{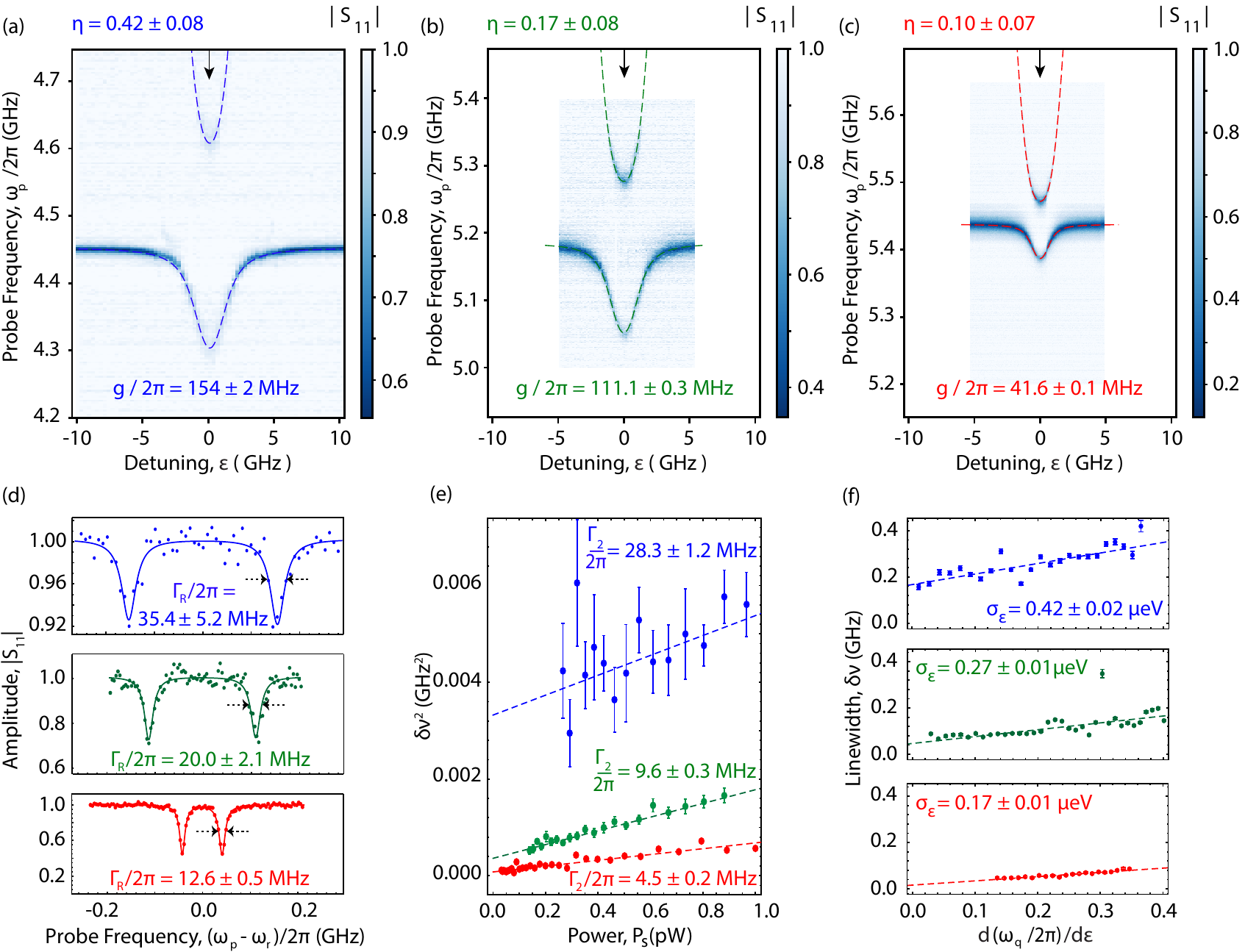}
	\caption{
		The dependence of the coupling strength $g$ and DQD coherence rates $\Gamma_{\rm{R}}$ and $\Gamma_2$ for DQD configurations with dipole strengths $\eta=0.42, \, 0.17,\,0.10$. 
		(a-c) Resonator reflectance amplitude $|S_{11}|$ versus DQD detuning $\detuning$ for three representative values of the dipole strength $\eta \sim 0.42 \pm 0.08$ (blue), $\eta \sim 0.17 \pm 0.08$ (green) and $\eta \sim 0.10 \pm 0.07$ (red) 
		[corresponding to the DQD charge stability diagrams in Fig.~\ref{fig2:four_stability_diagramss}(c), (b) and (a), respectively].
		(d) Resonator amplitude response $|S_{11}|$ (dots) \emph{vs.} probe frequency $\omega_\mathrm{p}/2\pi$ at $\detuning=0$ [see black arrow in the panels (a-c)], displaying well-resolved vacuum Rabi mode splittings. 
		The solid line is a fit to the sum of two Lorentzian lines. The quoted $\Gamma_{\rm{R}}$ is an average of the two linewidths.
		(e) Squared qubit linewidth $\delta\nu_{\rm{q}}^2$ (dots)
		\emph{vs.} spectroscopy drive power $P_s$, measured via two-tones spectroscopy \cite{Stockklauser2017}. The dashed lines are linear fits. The zero-power linewidths $\Gamma_2$ are given in the panel.
		(f) Qubit linewidth $\delta\nu_{\rm{q}}$ (dots) \emph{vs.} $d\omega_\mathrm{q}/d\detuning$ extracted from two-tones spectroscopy \cite{Stockklauser2017}. The dashed lines are linear fits. Their slopes define $\sigma_{\detuning}$ according to Eq.~\eqref{eq:decoherence}.
		}
	\label{fig3:Display_porperties}
\end{figure*}

In this first set of experiments, we employ a SQUID array resonator [see Figs.~\ref{fig1:SampleAndCircuit}(a,c)] with an estimated impedance $Z^{\mathrm{Sq}}_r=\sqrt{L_{\mathrm{Sq}}/C_{\mathrm{Sq}}}\sim1$~k$\Omega$. 
Similar high impedance resonators have been previously shown to enable the strong coupling regime between a DQD and microwave photons~\cite{Stockklauser2017}. 
A magnetic flux, applied via a superconducting coil mounted on the sample box, is used to tune the resonator in the frequency range $\omega_r/2\pi \sim [4.2,5.7] \, \rm{GHz}$ (see Tab.~\ref{tab1:para_fig3_large}). 
The internal resonator dissipation $\kappa_\mathrm{int}$ and coupling to the microwave feedline $\kappa_\mathrm{ext}$ change with the resonator frequency, as shown in Fig.~\ref{figKappas}(c) in the Appendix 
\footnote{The total dissipation of the SQUID array resonator is not constant as a function of the resonator frequency due to the presence of standing waves in its microwave feedline.}.

The DQD response to the gate voltages is characterized by charge stability diagrams \cite{vanderwiel2002} which we measure by recording the amplitude and phase response of the reflectance of the resonator~\cite{Frey2012}. 
From those diagrams, we extract the charging energies and capacitances of the DQD. 
In Figs.~\ref{fig2:four_stability_diagramss}(c-f) we present four typical examples of DQD charge stability diagrams realized within the same device by \emph{in-situ} tuning the voltages on the four gates defining the DQD [Fig.~\ref{fig1:SampleAndCircuit}(e)]. 
The differences between the four configurations lie mainly in different voltages applied to the gates T and CG [cf. Fig.~\ref{fig1:SampleAndCircuit}(e)], which control the interdot tunnel barrier, and are listed in Table~\ref{table:voltages} in App.~\ref{sec:appendix_determination_of_Cs}.
For ease of comparison, the axes ranges are identical for the four panels of Fig.~\ref{fig2:four_stability_diagramss}.
We stress again that these four different configurations present similar interdot tunneling amplitudes $\tunneling$ despite the different gate voltage values.

Comparing the four DQD configurations shown in Figs.~\ref{fig2:four_stability_diagramss}(c-f), we notice that the average spacing between the DQD triple points \cite{vanderwiel2002} 
[maximal in Fig.~\ref{fig2:four_stability_diagramss}(c)] decreases relative to the spacing between two consecutive QD-reservoirs charge transitions [dashed lines in Fig.~\ref{fig2:four_stability_diagramss}(c-f)].
This variation can be interpreted as a net change of the contribution of the interdot coupling capacitance ($\capMut$) to the total capacitance of the individual QDs ($\capTotal{1}$ and $\capTotal{2}$) \cite{vanderwiel2002}.
This translates into the dipole strength $\eta$, Eq.~\eqref{eq:suppression}, 
covering the interval $[0.1,\,0.7]$ in our experiments. 
Both $C_\mathrm{m}/C_\mathrm{\Sigma}$ and $\eta$ can be determined from the charge stability diagrams by considering the arrows indicated in the schematics in Figs.~\ref{fig2:four_stability_diagramss}(a,b). The red arrow represents the distance of two adjacent DQD triple points and the blue arrow connects two consecutive electron transitions with the leads. As derived in Appendix~\ref{sec:appendix_determination_of_Cs}, in the simplified case of symmetric quantum dots, $C_{\Sigma,1}=C_{\Sigma,2}=C_\Sigma$, and neglecting gate-cross capacitances, one finds $C_\mathrm{m}/C_\Sigma=\Delta V_\mathrm{m}/(\Delta V_\mathrm{g}-\Delta V_\mathrm{m})$ and $\eta=1 - \frac{2\Delta V_{\rm m}}{\Delta V_{\rm g}}$. $\Delta V_\mathrm{m}$ ($\Delta V_\mathrm{g}$) represents the length of the red (blue) arrow in Figs.~\ref{fig2:four_stability_diagramss}(a,b).
Furthermore, we emphasize that this striking change of the DQD interdot capacitance is obtained while keeping the interdot tunneling rate in the range $4\,\rm{GHz}<\tunneling/h < 6\,\rm{GHz}$. The ability to control $\tunneling$ and $\eta$ independently allows us to probe the interaction with the resonator in both resonant and dispersive regimes. 

In this section, we study a total of eleven different DQD configurations, whose extracted parameters are summarized in Tab.~\ref{tab1:para_fig3_large}.
For three of these configurations, we present in Fig.~\ref{fig3:Display_porperties} 
the hybridized qubit-resonator energy spectrum [see  Fig.~\ref{fig3:Display_porperties}(a-d)],
a measurement of the intrinsic DQD charge qubit linewidth [see  Fig.~\ref{fig3:Display_porperties}(e)],  
and a measurement of the root-mean-square amplitude of the detuning noise $\sigma_{\detuning}$ defined in Eq.~\eqref{eq:decoherence} [see  Fig.~\ref{fig3:Display_porperties}(f)].
With the exception of the measurements reported in panel (f), all the data plotted in Fig.~\ref{fig3:Display_porperties} were taken at the charge degeneracy ($\detuning=0$). 

We now discuss the three independent measurements reported in Fig.~\ref{fig3:Display_porperties}.
In Fig.~\ref{fig3:Display_porperties}(a-c) we show three examples of hybridized spectra in the strong coupling regime for different dipole strengths. 
The DQD stability diagrams of the three configurations in Fig.~\ref{fig3:Display_porperties}(a,b,c) are shown in panels (e,d,c) of Fig.~\ref{fig2:four_stability_diagramss}, respectively, in corresponding colors.
We tune the DQD gate voltages and the SQUID array resonant frequency to reach the resonance condition 
$\omega_\mathrm{q} =\omega_\mathrm{r}$ at approximately zero detuning $\detuning$.
Varying the DQD detuning, we observe the characteristic shifts in the dispersive regime and clear indications of an avoided crossing~\cite{Mi2017,Stockklauser2017} at resonance.
We analyze the hybridized spectrum and extract the coupling strength $\qubitResonatorCoupling$, resonator resonance frequency $\resonatorAngularFrequency /2\pi$, and DQD tunneling amplitude $\tunneling$ 
by fitting the observed resonances to the spectrum extracted from the simulation of the system Hamiltonian (see Appendix~\ref{sec:masterEq} for details). The latter is plotted by dashed lines in Figs.~\ref{fig3:Display_porperties}(a-c).

When comparing these three configurations, we take note of a correlation between the coupling strength $\qubitResonatorCoupling$ and the visibility of the reflected signal (Rabi modes splitting) around the avoided crossing.
To help visualizing this correlation, Fig.~\ref{fig3:Display_porperties}(d) shows the linecuts at the resonance [detuning indicated by black arrows in Figs.~\ref{fig3:Display_porperties}(a-c)).
Furthermore, when increasing $\eta$ we observe a distinct increase of the linewidth of the Rabi modes $[\Gamma_{\rm{R}} \sim (\kappa_{\rm{ext}} + \kappa_{\rm{int}})/2 + \Gamma_2]$, 
extracted by fitting the data to a sum of two Lorentzian lines [see solid line in Fig.~\ref{fig3:Display_porperties}(d)], and a clear reduction in the depth of the two Lorentian (compare the y-axis for the three panles of Fig.~\ref{fig3:Display_porperties}(d)).
This suggests that the dipole strength also has a strong influence on the system decoherence.

We investigate in more detail the correlations between the measured dipole strength $\eta$ andù
the observed coherence of the charge qubit. 
Using two-tone spectroscopy~\cite{Schuster2005,Stockklauser2017}, we measure the intrinsic qubit linewidth at the detuning sweet spot ($\detuning=0$)
and its sensitivity to the noise in the detuning parameter induced by the charge noise of the electromagnetic environment of the DQD
\footnote{
These measurements are implemented by changing the resonator frequency for performing these experiments in the dispersive regime, ensuring a negligible contribution of the Purcell induced decay ($\Gamma_\mathrm{purcell}/2\pi<0.05$~MHz). Also, we drive the resonator weakly so
  that its population on average is $<1$ photon. The reflected signal is then
  amplified via a Josephson parametric amplifier with a gain of $\sim
  18$~dB.}.
Measureing the power dependence of the qubit linewidth we extract the zero power linewidth ($P_S\rightarrow0$) [see Fig.~\ref{fig3:Display_porperties}(e)], which allows to determine the intrinsic DQD charge decoherence rate $\Gamma_2$~\cite{Schuster2005,Stockklauser2017}.
In this experiment, we reached a DQD linewidth as low as $\sim4.5\pm 0.2$ MHz for a configuration with $\eta=0.10 \pm 0.07$. 
In contrast, by \textit{in-situ} tuning to a configuration with $\eta=0.71 \pm 0.03$, the DQD charge qubit linewidth increases by a factor of eight.

\begin{figure*}[!t]
\includegraphics[width=\textwidth]{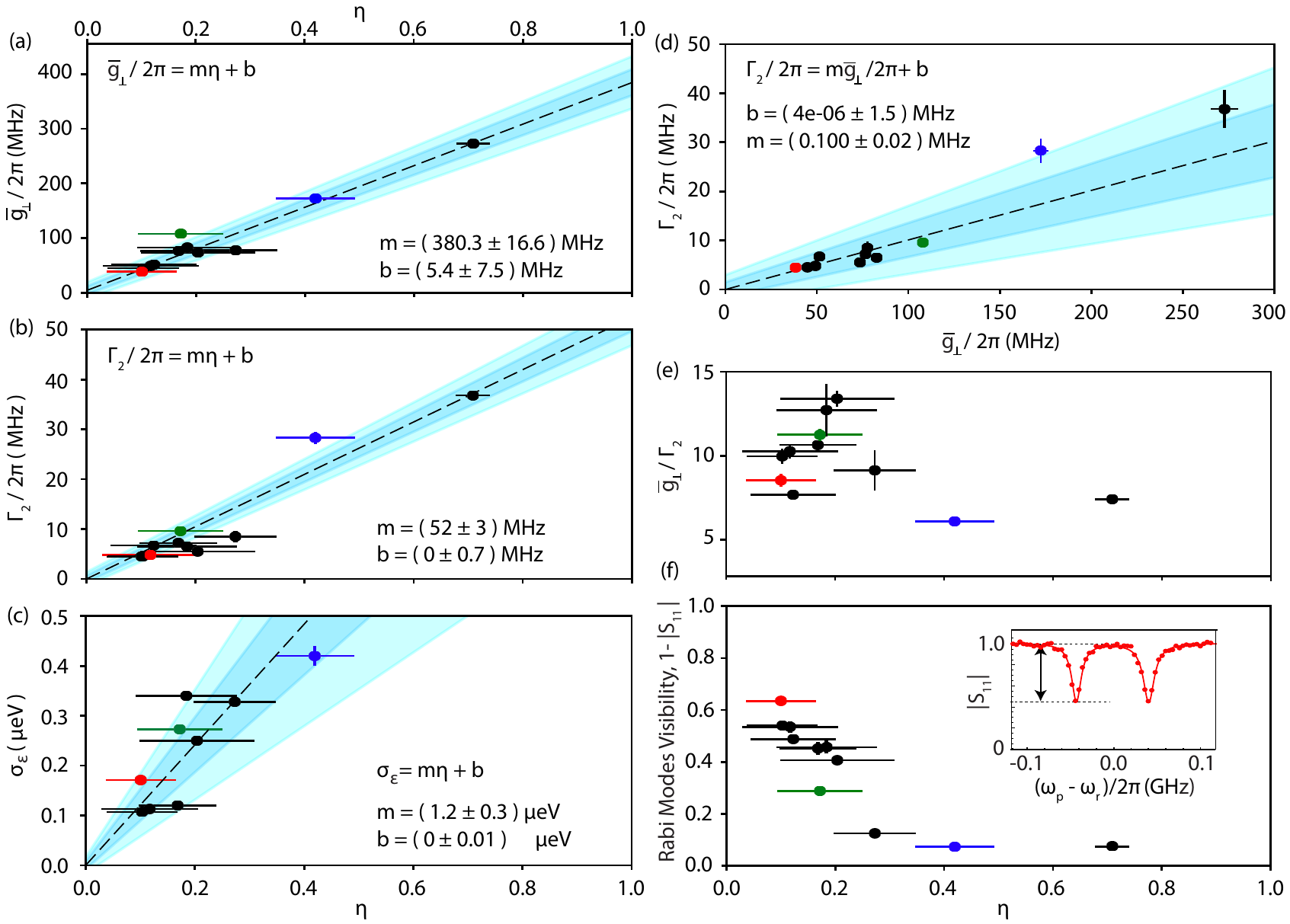}
	\caption{
		Coupling strength and decoherence parameters extracted for eleven DQD gate bias voltage configurations.
		(a) Normalized coupling $\normCoupling$ [see Eq.~\eqref{eq:normalization}] of the DQD charge qubit to the resonator
		% normalized for $\detuning =0$ and $\Delta = 5\, \rm{GHz}$, $\normCoupling = \qubitResonatorCoupling \frac{\tunneling}{\resonatorAngularFrequency} \frac{5\, \rm{GHz}}{\resonatorAngularFrequency /2\pi}$ 
		\emph{vs.} the dipole strength $\eta$. 
		(b) Qubit linewidth $\Gamma_2$ \emph{vs.} $\eta$. The linewidth is extracted as in Fig.~\ref{fig3:Display_porperties}(e). 
		(c) Effective detuning noise of the DQD charge qubit $\sigma_{\detuning}$ \emph{vs.} $\eta$, obtained as in Fig.~\ref{fig3:Display_porperties}(f). 
		For the two configurations in correspondence of $\eta=0.123$ and $0.709$ we could not extract $\sigma_{\detuning}$ due to either spurious resonances and enhanced sensitivity to detuning noise, respectively.
		(d) DQD linewidth $\Gamma_2$ \emph{vs.} the normalized coupling. 
		The data in (a)-(d) were fitted to a linear model plotted as dashed lines and the fit parameters are stated in the panels.
		The dark [light] blue area represents the one-[two-]sigma confidence interval.
		(e) The quality factor $\normCoupling/\Gamma_2$ \emph{vs.} $\eta$.
		(f) Visibility of the vacuum Rabi modes (at resonance) $(1-\lvert S_{11}\rvert )= 2\kappa_\mathrm{ext}/(\kappa_\mathrm{ext}+\kappa_\mathrm{int}+2\Gamma_2)$ \emph{vs.} $\eta$. 
		The insert shows an example of a vacuum Rabi mode splitting with the black arrow indicating the visibility of a Rabi mode at the resonance.
		}
	\label{fig3:Plot_parameters}
\end{figure*}

At $\detuning=0$, the charge qubit is in first-order insensitive to charge noise since $\partial \qubitAngularFrequency / \partial \detuning = 0$. 
Measuring the dependence of the qubit linewidth vs.\ the detuning $\detuning$, we extract the detuning noise $\sigma_{\detuning}$ according to Eq.~\ref{eq:decoherence} [see Fig.~\ref{fig3:Display_porperties}(f) and Fig.~\ref{fig3:Plot_parameters}(c) and also Refs.~\cite{Scarlino2017b,Thorgrimsson2017}].
The extraction of $\sigma_{\detuning}$ in two-tone spectroscopy was performed at a larger resonator read-out power explaining the lower error bars on the extracted linewidths and the higher value of qubit linewidth at $\detuning = 0$ compared to Fig.~\ref{fig3:Display_porperties}(e). 
We notice that $\sigma_{\detuning}$ clearly decreases for lower $\eta$.

The measurements presented in Fig.~\ref{fig3:Display_porperties}, indicate that increasing the capacitance ratio $C_{\rm{m}}/C_{\Sigma}$
reduces the resonator-DQD coupling strength $\qubitResonatorCoupling$ [Fig.~\ref{fig3:Display_porperties}(d)], the qubit decoherence $\Gamma_2 \equiv \delta \nu (P_{\rm{S}} \rightarrow 0)$ [Fig.~\ref{fig3:Display_porperties}(e)], and the sensitivity of qubit energy to detuning noise
[Fig.~\ref{fig3:Display_porperties}(f)]~\cite{You2005a}. 
This reduced sensitivity of the DQD to charge noise is engineered through a large mutual capacitance of strongly coupled QDs.

We summarize results of similar measurements for all eleven investigated DQD configurations in Fig.~\ref{fig3:Plot_parameters}.
In order to systematically compare the coupling strength $\qubitResonatorCoupling$ of the different configurations, we normalize it to \footnote{From unpublished data, reported in Fig.~\ref
  {fig_glinear}(a) in Appendix~\ref {secrenormalization}, which will be the
  topic of another manuscript, we observe an unexpected linear dependence of
  the coupling strength $g$ on the resonator frequency $\omega _r(=\omega _q)$
  measured at the sweet spot, $\epsilon =0$. We take this into account by
  defining the normalized coupling $\protect \bar g_{\perp }$, instead of the
  expected $\protect \bar g_{\perp }^{'} \propto \protect \sqrt {\omega _r}$
  dependence.}
\begin{equation}
\normCoupling = \qubitResonatorCoupling(\resonatorAngularFrequency /2\pi = \tunneling = 5 \,\rm{GHz}) = \qubitResonatorCoupling \frac{\tunneling}{\resonatorAngularFrequency} \frac{5\, \rm{GHz}}{\resonatorAngularFrequency /2\pi}.
\label{eq:normalization}
\end{equation}
The normalization aims to systematically account for the small differences in the resonator frequency/inductance and in DQD tunneling amplitude $\tunneling$~\cite{Stockklauser2017}
at which the experiments exploring the different bias conditions were performed at
(see Table~\ref{tab1:para_fig3_large} and Appendix~\ref{secrenormalization}).
The normalized coupling strength ranges from 41.6 MHz to 250.6 MHz.
The dependence of the normalised coupling $\normCoupling$ on $\eta$ agrees with the linear relation [see Fig.~\ref{fig3:Plot_parameters}(a)] derived as Eq.~\eqref{eq:coupling}. 

A similar dependence on $\eta$ is also observed for the DQD decoherence $\Gamma_2$
[Fig.~\ref{fig3:Plot_parameters}(b)] and detuning noise $\sigma_{\detuning}$ [Fig.~\ref{fig3:Plot_parameters}(c)], as modeled by Eq.~\eqref{eq:decoherence}.
In order to display the linear relation between coupling strength $\normCoupling$ and DQD decoherence $\Gamma_2$, we plot them on the two axes of Fig.~\ref{fig3:Plot_parameters}(d). 
The scattered $(\Gamma_2, \normCoupling)$ data lies within the 3$\sigma$ confidence interval of the linear fit.
This proportionality relation is additionally highlighted by inspecting the 
quality factor of the resonator-qubit hybrid system $Q=\normCoupling / \Gamma_2$~\cite{Cottet2017b}.
In Fig.~\ref{fig3:Plot_parameters}(e) we observe that $Q$ does not show strong dependence on the dipole strength $\eta$, but it is scattered around the a mean value 9.7 with a standard deviation of 2.2, indicating that the coherence of the system is likely dominated by dephasing due to singular charge noise (see Sec.I).

For a circuit QED architecture realized with semiconductor QDs and superconducting resonators, the strong coupling regime has been reached only recently~\cite{Mi2017,Stockklauser2017}. 
Intrinsic limitations are the usually high decoherence rate of the orbital-charge degree of freedom and the small electric dipole moment of electrons confined in QDs.
The high qubit decoherence usually implies low visibility of the vacuum  Rabi mode splitting, even if the strong coupling is reached \cite{Stockklauser2017}. 
In Appendix \ref{sec:visibility}, we derive an expression for the visibility of the vacuum-Rabi mode splitting for a single port resonator coupled to a DQD and tested in reflection. For the case where DQD and resonator are tuned into resonance, we find $(1-|S_{11}|) = 2 \kappa_\mathrm{ext}/(\kappa_\mathrm{ext} + \kappa_\mathrm{int} + 2\Gamma_2)$. 
This estimated visibility is plotted in Fig.~\ref{fig3:Plot_parameters}(f) for the different DQD configurations explored in this study.
% (the values of $\kappa_\mathrm{ext}$ and $\kappa_\mathrm{int}$ for the different configurations are summarized in Fig.~\ref{figKappas}(c) in Appendix). 
When tuning the DQD into a configuration where the interdot capacitance is the dominant contribution ($\eta\rightarrow 0$), the Rabi mode splitting visibility is considerably increased despite a reduction in the coupling strength.
Furthermore, it is instructive to consider the system cooperativity, defined as $C = \normCoupling^2/[\Gamma_2 (\kappa_{\rm{ext}}+\kappa_{\rm{int}})]$, representing a dimensionless measure of the light/matter interaction strength in our hybrid system.
%and compares to what already achieved for similar devices deployed in previous experiments. 
As reported in Appendix \ref{Figures of Merit} [see Fig.~\ref{figS2}(a)], in this work we have achieved $C>100$ by making use of the described tuning strategy for the DQD electric dipolar energy. This represents
%the limits for the cooperativy achieved for the semiconductor QD-resonator systems above 100, representing 
the highest cooperativity reported so far for hybrid QD-resonator systems (see Ref.\cite{Cottet2017b} for a comparison), even when comparing to hybrid spin-photon systems. 

To summarize the results of this set of experiments on the first sample realized using a SQUID array for the resonator, we observe a striking and clear dependence of the DQD-resonator coupling strength, DQD charge decoherence rate, and DQD detuning noise on the dipole strength, parametrized by $\eta$, as defined in Eq.~\ref{eq:suppression}.
The characterization of different DQD configurations, realized by changing \textit{in-situ} the voltages applied to the DQD depletion gates over an extensive voltage range, demonstrates the possibility to reduce the charge qubit decoherence rate down to less than~5~MHz, thanks to the reduced DQD electric dipole strength.
The improved charge coherence allows to considerably increase the visibility of the charge qubit-resonator Rabi vacuum mode splitting at small coupling strengths with good coherence.

\section{Ultrastrong coupling with a junction array resonator \label{sec:USC}}
\begin{figure*}[!t]
\includegraphics[width=\textwidth]{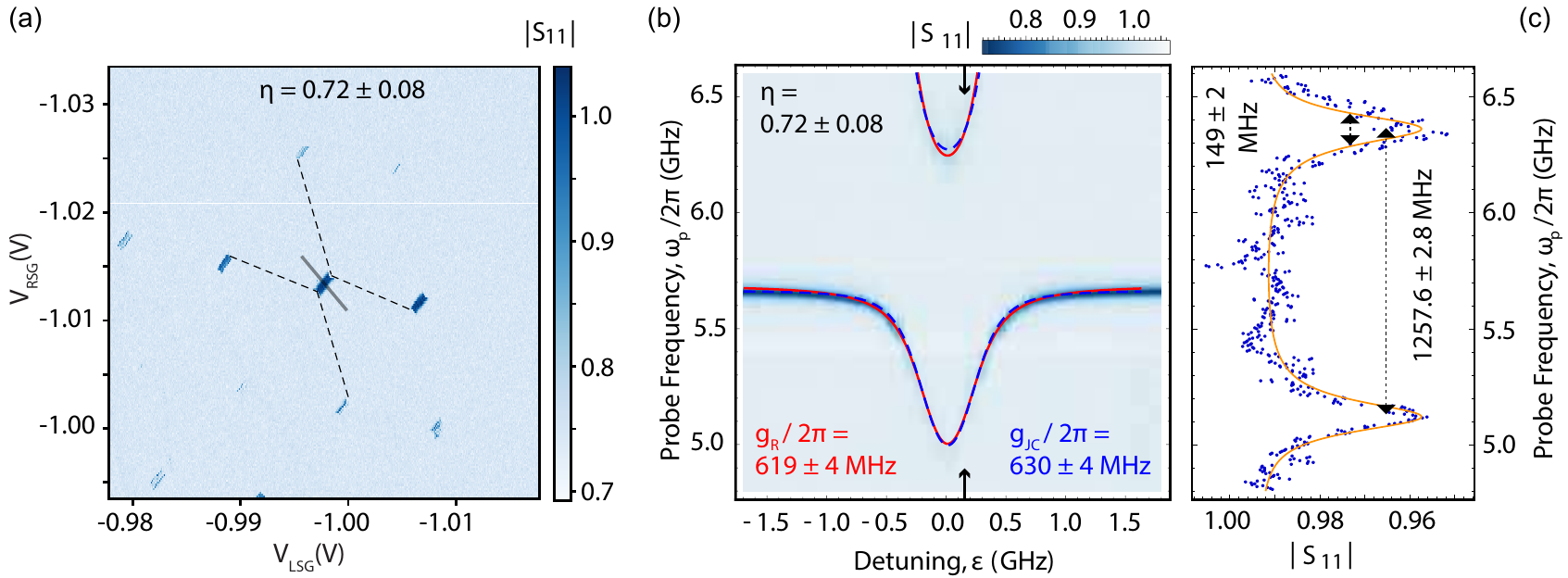}
	\caption{
		Investigation of a bias configuration approaching the ultra-strong coupling regime for a DQD coupled to a JJ array resonator. 
		(a) Charge stability diagram of the DQD measured by monitoring the change in resonator reflectance amplitude $|S_{11}|$ for the extracted dipole strength $\eta = 0.72 \pm 0.08$.
		(b) Resonator amplitude response $|S_{11}|$ taken by varying the DQD detuning $\detuning$ along the grey line indicated in panle (a) by applying appropriately chosen voltages to the two side gates.
		Red (blue) line represents a fit to the Rabi (JC) model (see Appendix \ref{sec:masterEq}).
% from which we extract 
%		$\tunneling_{\rm{R}}/2\pi = 5.547 \pm 0.006$ GHz ($\tunneling_{\rm{JC}}/2\pi= 5.561 \pm 0.006$ GHz),
%		$\omega_{r,\rm{R}}/2\pi \sim 5.6580 \pm 0.0005~\rm{GHz}$ ($\omega_{r,\rm{JC}}/2\pi \sim 5.6594 \pm 0.0005~\rm{GHz}$)  
%		and $g_{\rm{R}}/2\pi=350\pm3$ MHz ($g_{\rm{JC}}/2\pi=351\pm3$ MHz). See also Fig.~\ref{figS4}(a).
(c) Measured resonator reflectance $|S_{11}|$ (dots) \emph{vs.} probe frequency $\omega_{\rm{p}}$ extrated at resonance for $\detuning=0.15\,\rm{GHz}$ (black arrows in panel (b)), displaying a vacuum Rabi mode splitting.
%		(c) Linecut representing $|S_{11}|(\omega_{\rm{p}}/2 \pi)$ taken along the black arrows in (b). 
		The orange line represents a fit to a Rabi master equation model.
%		from which we extract $g/2\pi \sim 373.4 \pm 0.6 ~\rm{MHz}$ [$g/2\pi \sim 628.8 \pm 1.4 ~\rm{MHz}$] and $\Gamma_{2}/2\pi \sim 56.3 \pm0.2\, \rm{MHz}$ [$\Gamma_{2}/2\pi \sim 149 \pm\, 2 \rm{MHz}$] in correspondence of $\eta \sim 0.50 \pm 0.14$ [$\eta \sim 0.72 \pm 0.08$] 
		The JJ array resonator losses are $\kappa_{int}/2\pi=19.5 \,\rm{MHz}$ and $\kappa_{ext}/2\pi=5.7\pm0.1\,\rm{MHz}$.
% from which we extract 
%		$\tunneling_{\rm{R}}/2\pi = 5.581 \pm 0.008$ GHz ($\tunneling_{\rm{JC}}/2\pi= 5.612 \pm 0.007$ GHz),
%		$\omega_{r,\rm{R}}/2\pi \sim 5.6612 \pm 0.0003~\rm{GHz}$ ($\omega_{r,\rm{JC}}/2\pi \sim 5.6635 \pm 0.0002~\rm{GHz}$)  
%		and $g_{\rm{R}}/2\pi=619\pm4$ MHz ($g_{\rm{JC}}/2\pi=629\pm4$ MHz). See also Fig.~\ref{figS4}(b).
	}
	\label{fig6:Ultrastorngcoupling}
\end{figure*}

In Sec.~\ref{sec:Coherence}, we have investigated the possibility to \emph{in-situ} tune the DQD dipolar coupling energy. We have explored the trade-off between the qubit-resonator coupling and the DQD charge decoherence rate.
In this section, we show that the same strategy allows us to approach the ultrastrong coupling regime. 
With this goal in mind, we have realized a second device. It is similar to the first one but 
for the superconducting microwave resonator, which is now formed by a more compact Josephson junction (JJ) array replacing the SQUID array \cite{Masluk2012}.
Replacing SQUIDs with single Josephson junctions in the array makes the resonator fixed in frequency, which reduces your flexibility on tuning parameters. 
On the other hand, as illustrated in Fig.~\ref{fig1:SampleAndCircuit}(b) and explained in Appendix \ref{sec:resonators}, this change reduces the dimensions of the array unit. We thus achieve a higher Josephson inductance with a shorter array: the length of the JJ resonator 
is $\sim70 ~\mu \rm{m}$, instead of $\sim 250 ~\mu m$ for the SQUID array [compare Fig.~\ref{fig1:SampleAndCircuit}(a,b) and Fig.~\ref{fig1:SampleAndCircuit}(c,f)]. The JJ array resonator has a lower stray capacitance to ground to $C_{\rm{gnd}}^{\rm{JJ}} \sim 5 \, \rm{fF}$, with a total
inductance of $L_{\rm{tot}}^{\rm{JJ}} \sim 100 \, \rm{nH}$ and, in turn, a resonator impedance 
$Z_{\rm{r}}^{\rm{JJ}} \sim 4 \, \rm{k}\Omega$.
Parameters of the SQUID and JJ arrays are compared in Tab.~\ref{table:nonlin}.

Replacing SQUIDs with single Josephson junctions in the array makes the resonator fixed in frequency, which reduces your flexibility on tuning parameters. 

Aiming at realizing the ultrastrong coupling regime with semiconductor quantum dots, we investigate a DQD configuration
corresponding to the largest dipole strenght that we were able to achieve with this second device, having a dipole strength $\eta \approx 0.72$.
As discussed in Sec.~\ref{sec:Coherence}, we detect the amplitude and phase of the signal reflected off the resonator.
We configure the DQD tunneling amplitude close to $\tunneling / h \sim \omega_{\rm{r}}/2\pi$ and change the DQD detuning. 
Upon bringing the qubit energy into resonance with the resonator, $\omega_{\rm{q}} \sim \omega_{\rm{r}}$, 
a clear avoided crossing is observed  in the resonator reflectance [see Fig.~\ref{fig6:Ultrastorngcoupling}(b), and Fig.~\ref{figS4}(b)]. This is a sign of reaching the strong coupling regime. 

The data are in excellent agreement with the spectrum of the hybridized system numerically calculated 
using 
$\qubitResonatorCoupling$, $\resonatorAngularFrequency$ and $\tunneling$ as adjustable parameters.
We fit a Rabi (red solid line) and a Jaynes–Cummings (blue dashed line) model to this data and present the results in Fig.~\ref{fig6:Ultrastorngcoupling}(b). We extracted a coupling strength $g_{\rm{R}}/2\pi \sim 620 \pm 2 ~\rm{MHz} $ ($g_{\rm{JC}}/2\pi  \sim 637 \pm 2 ~\rm{MHz} $) from which we can estimate a $g_{\rm{R,JC}}/\omega_{\rm{r}} \sim 0.11 \, \pm \, 0.01$, reaching the ultrastrong coupling regime~\cite{Niemczyk2010, Kockum2018a, FornDiaz2019}.
The discrepancy between the values obtained from the Rabi and JC fits is due to the onset of the USC regime 
\footnote{Fig.~\ref {fig6:Ultrastorngcoupling_top} in Appendix~\ref
  {Extra data} reports the same analysis performed on a second DQD
  configuration characterized by $\eta \sim 0.5$. There we obtain comparable
  results from fits to the Rabi and JC models.}.
The resonator reflectance $|S_{11}|$ \emph{vs.} probe frequency $\omega_{\rm{p}}$  at the DQD-resonator detuning value indicated by the black arrow in Fig.~\ref{fig6:Ultrastorngcoupling}(b) (resonant condition) is shown in Fig.~\ref{fig6:Ultrastorngcoupling}(c).
By fitting a master equation model [see solid orange line in Fig.~\ref{fig6:Ultrastorngcoupling}(c)] to the measured $|S_{11}|$ we extract a DQD decoherence of $\Gamma_{2}/2\pi \sim 149 \pm 2 \, \rm{MHz}$
% $\Gamma_2 \sim 50 \pm ~\rm{MHz}$ 
and a Rabi mode splitting of $2\qubitResonatorCoupling/2\pi \sim 1258 \pm 3~ \rm{MHz}$. 
Resolving the two dips of the vacuum Rabi mode splitting indicates that the system is still in the strong coupling regime despite the extra decoherence introduced by the large DQD electric dipole strength.

\section{Conclusions\label{sec:Conclusion}}
We have realized two hybrid devices with which
we have studied charge configurations at the two extremes
of the explored tunable DQD electric dipole strength. 
We have demonstrated the systematic control of the DQD electric dipole strength, allowing us to explore a broad range of different regimes in the same device.
In particular, we have demonstrated that it is possible to decrease the electric dipolar coupling energy of the DQD by tuning it into a configuration in which the interdot mutual capacitance $C_{\rm{m}}$ becomes the dominant contribution of the total DQD capacitance. In such a configuration, 
the small dipole strength ($\eta \rightarrow 0$)
reduces both the DQD coupling to the resonator and its decoherence rate, 
down to a 
$g/ 2 \pi\sim$~40 MHz and
$\Gamma_2/2\pi<$~5 MHz, respectively.

We have made use of the control of the DQD dipole strength reported here to reduce the decoherence rate of DQD devices used in some of our previous works. It has led to the observation of a DQD qubit linewidth down to $\Gamma_2/2\pi\sim3$~MHz in a similar device~\cite{Scarlino2017b,Scarlino2018}.
These decoherence rates are well below values reported typically for semiconductor charge qubits, usually observed to be above hundreds of~MHz or even up to several GHz~\cite{Petersson2010,Basset2013,Stockklauser2017}.
The possibility to achieve these remarkably low decoherence rates for a DQD charge qubit enabled the realization of time-resolved dispersive read-out~\cite{Scarlino2017b}, and distant qubit-qubit interaction mediated by virtual microwave photons~\cite{Scarlino2018,vanWoerkom2018}.

Here, we have provided a detailed explanation and a method to engineer these low charge decoherence values by easily modifying the contribution of the interdot capacitance $C_m$ to the total QD capacitance, which we can easily assess and tune by exploring the DQD charge stability diagram.
Furthermore, this experiment sheds new light on the puzzling observation reported by different experiments on QD-resonator hybrid system \cite{Scarlino2017b,Ibberson2020} which observed that $g$ and $\Gamma_2$ can vary considerably within the same device configured in different regions of the DQD charge stability diagrams.

In addition, we show that by using the same tuning strategy of the DQD confinement potential, but striving to maximize the DQD electric dipolar coupling energy, we can considerably increase the DQD-resonator coupling strength. This is achieved by configuring the DQD gates voltages to minimize the interdot capacitance $C_{m}$.
To further increase the coupling strength, we implemented a more compact Josephson junction array resonator with reduced stray capacitance respect to a SQUID array resonator. This results in a $\sim4\,\rm{k\Omega}$ resonator impedance.
The JJ array resonator enabled a maximum coupling of $\qubitResonatorCoupling/2\pi \sim 630 \, \rm{MHz}$ for a fundamental mode resonator frequency of $\resonatorAngularFrequency/2\pi \sim5.6\,\rm{GHz}$.
In this way, we realize the Ultra Strong Coupling regime
between electrons hosted in a semiconductor DQD and a microwave resonator.
By increasing the resonator impedance even further and by defining DQDs in shallower 2DEGs, or in semiconductor nanowires and Si-CMOS devices, where a higher gate lever-arm (up to 0.75 in \cite{deJong2019}) has been demonstrated for QDs, it may well be possible to achieve $\qubitResonatorCoupling /\omega_{\rm{r}} \sim 0.4-0.5$. This could enable more advanced investigations of the effects of the breakdown of the rotating-wave-approximation in this class of light-matter hybrid devices~\cite{Niemczyk2010, Kockum2018a, FornDiaz2019}.

Recent experiments with holes confined in 2D-Ge heterostructures have reported effective charge/gate noise lower by a factor of 2-4 with respect to Si and GaAs 2DEG systems~\cite{Lodari2020}, estimated by recording the current fluctuations of a charge detector over long waiting times.
Applying the dipole strength tuning strategy described in this manuscript to holes confined in QDs defined in 2D-Ge systems may enact a substantial improvement in the coherence properties of the charge degree of freedom.
This could enable a more clear study of the ultrastrong coupling physics in the $\eta \rightarrow 1$ limit and the potential to achieve $\mu s$ coherence time for a DQD charge qubit in the $\eta \rightarrow 0$ limit.

Understanding and improving the coherence and control of the electron/hole charge degree of freedom in semiconductor systems is of paramount importance also for future spin qubit applications, especially for systems where the spin is strongly hybridized with the orbital degree via a high real \cite{froningultrafast} or artificial spin-orbit field \cite{kawakami2014}.
We anticipate that these findings will be of great significance for state-of-the-art charge and/or spin qubits as well as any hybrid designs, which are all limited by electrical noise.

\begin{acknowledgements}
	We acknowledge Udson Mendes, Christian Andersen, Mihai Gabureac, Theo Walter, Johannes Heinsoo, Philipp Kurpiers, for the useful discussion. We thank Alexandre Blais for valuable feedback on the manuscript.
	This work was supported by the Swiss National Science Foundation through the National Center of Competence in Research (NCCR) Quantum Science and Technology, 
	the project Elements for Quantum Information Processing with Semiconductor/Superconductor Hybrids (EQUIPS) and by ETH Zurich.
\end{acknowledgements}

\begingroup
\bibliographystyle{apsrev4-1}
\def\refname{References}

\def\bibname{References}

\endgroup

%\bibliographystyle{apsrev4-1}
%\bibliography{JabRef/QudevRefDB2,JabRef/QudevRefDB_PS,JabRef/TheoryRefs}
%\bibliography{Refs/QudevRefDB2,Refs/QudevRefDB_PS,Refs/TheoryRefs}
%\begin{thebibliography}{}
%   \input{In_situ_tuning_of_the_electric_dipole_strength_DQD.bbll}
%\end{thebibliography}
%\bibliography{In_situ_tuning_of_the_electric_dipole_strength_DQD.bbl}

%\widetext

\clearpage

\newpage

\appendix

%\begin{widetext}

%\section{Appendix}\
\beginsupplement

\section{Experimental determination of the dipole strength $\eta$ \label{app:cap_model}}
\label{sec:appendix_determination_of_Cs}

\begin{table*}[ht]
\centering
\begin{tabular}{|c|c|c|c|c|c|c|c|c|c|}
  \hline
  % after \\: \hline or \cline{col1-col2} \cline{col3-col4} ...
  index & $C_{\Sigma,1}$~[fF] & $C_{\Sigma,2}$~[fF] & $C_{m}$~[fF] & $\eta$ & $\qubitResonatorCoupling /2\pi$~[MHz] & $\Gamma_2/2\pi$~[MHz] & $\sigma_{\detuning}$~[$\mu$eV] & $\tunneling/2\pi$~[MHz] & $\omega_r/2\pi$~[MHz] \\ \hline\hline
1 &  0.561$\pm$0.034 & 0.634$\pm$0.071 & 0.488$\pm$0.041 & 0.101$\pm$0.064 & 41.63$\pm$0.06 & 4.5$\pm$0.2 & 0.171$\pm$0.006 & 5420.8$\pm$0.2& 5437.0$\pm$0.1 \\ \hline
 2 & 0.433$\pm$0.037 & 0.474$\pm$0.061 & 0.358$\pm$0.041 & 0.117$\pm$0.088 & 54.9$\pm$0.1 & 4.8$\pm$0.2 & 0.113$\pm$0.009 & 5568.6$\pm$0.3& 5575.6$\pm$0.14\\ \hline
3 & 0.599$\pm$0.056 & 0.565$\pm$0.034 & 0.473$\pm$0.038 & 0.103$\pm$0.065 & 48.8$\pm$0.2 & 4.5$\pm$0.2 & 0.107$\pm$0.007 & 5435.1$\pm$0.5& 5578.6$\pm$0.11\\ \hline
 4 &  0.554$\pm$0.068 & 0.41$\pm$0.075 & 0.364$\pm$0.060 & 0.204$\pm$0.105 & 75.7$\pm$0.2 & 5.5$\pm$0.2 & 0.250$\pm$0.008 & 5137.4$\pm$0.4& 5117.6$\pm$0.14 \\ \hline
5&  0.656$\pm$0.065 & 0.70$\pm$0.053 &0.506$\pm$0.052 & 0.123$\pm$0.079 & 56.4$\pm$0.5 & 6.7$\pm$0.2 & - & 5482$\pm$3& 5578.4$\pm$0.4\\ \hline
6 &  0.611$\pm$0.053 & 0.54$\pm$0.058 & 0.443$\pm$0.046 & 0.168$\pm$0.071 & 86.3$\pm$0.2 & 7.2$\pm$0.2 & 0.120$\pm$0.007 & 5633.5$\pm$0.4& 5649.0$\pm$0.2\\ \hline
7 &  0.265$\pm$0.045 & 0.31$\pm$0.051 & 0.191$\pm$0.034 & 0.184$\pm$0.092 & 87.2$\pm$0.4 & 6.5$\pm$0.8 & 0.34$\pm$0.007 & 5276$\pm$1& 5283.7$\pm$0.6\\ \hline
8 &  0.333$\pm$0.031 & 0.27$\pm$0.041 & 0.250$\pm$0.026 & 0.172$\pm$0.078 & 111.1$\pm$0.3 & 9.6$\pm$0.3 & 0.273$\pm$0.005 & 5145$\pm$1& 5180.3$\pm$0.2\\ \hline
9 &  0.136$\pm$0.045 & 0.32$\pm$0.037 & 0.058$\pm$0.017 & 0.419$\pm$0.073 & 153.6$\pm$1.9 & 28.3$\pm$1.2 & 0.42$\pm$0.02 & 4453$\pm$4& 4440.9$\pm$0.3 \\ \hline
10 &  0.330$\pm$0.050 & 0.20$\pm$0.023 & 0.048$\pm$0.007 & 0.709$\pm$0.031 & 260.5$\pm$3.5 & 36.8$\pm$0.9 & - & 4772.7$\pm$9& 4745.5$\pm$0.9 \\ \hline
11 & 0.412$\pm$0.029 & 0.20$\pm$0.050 & 0.257$\pm$0.029 & 0.273$\pm$0.076 & 65.9$\pm$0.7 & 8.5$\pm$1.1 & 0.328$\pm$0.005 & 4243$\pm$2& 4271.6$\pm$0.2 \\ \hline
\end{tabular}
\caption{
%\cm{Can maybe go into appendix? Exact numbers do not seem important, especially as the relevant quantities $g, \eta$ are already summarized in Fig 4.}
	Extracted parameters for the eleven DQD configurations presented in Fig~\ref{fig3:Plot_parameters} in Sec.~\ref{sec:Coherence} of the main text.  
	}
\label{tab1:para_fig3_large}
\end{table*}

\begin{table*}[ht]
\centering % used for centering table
\begin{tabular}{|c| c| c| c| c|} % centered columns (4 columns)
\hline %inserts double horizontal lines
index & $V_{\rm{CG}}(mV)$ & $V_{\rm{SD}}(mV)$ & $V_{\rm{LS}}(mV)$ & $V_{\rm{RS}}(mV)$ \\ [0.5ex] % inserts table
%heading
\hline\hline % inserts single horizontal line
1&	-0.823&	-0.623&	-0.88132&	-0.946477273 \\ \hline
2&	-0.823&	-0.623&	-0.883236&	-0.937345455 \\ \hline
3&	-0.823&	-0.727&	-0.884445&	-0.789789091 \\ \hline
4&	-0.823&	-0.818&	-0.69147&	-0.751603636 \\ \hline
5&	-0.847&	-0.847&	-0.671525&	-0.6412 \\ \hline
6&	-0.882&	-0.882&	-0.60214&	-0.648681818 \\ \hline
7&	-0.936&	-0.936&	-0.79571&	-0.593763636 \\ \hline
8&	-0.982&	-0.982&	-0.576544&	-0.613915909 \\ \hline
9&	-1.04&	-1.04&	-0.473037&	-0.562018182 \\ \hline
10&	-1.05&	-1.05&	-0.49628&	-0.574921818 \\ \hline
11&	-1.03&	-1.03&	-0.525558&	-0.494352727 \\ [1ex] \hline
\end{tabular}
\caption{The DQD gate voltages for the eleven configurations investigated in Sec.~\ref{sec:Coherence} of the main text.} % title of Table
\label{table:voltages} % is used to refer this table in the text
\end{table*}

\begin{figure*}[!htb]
\includegraphics[width=\textwidth]{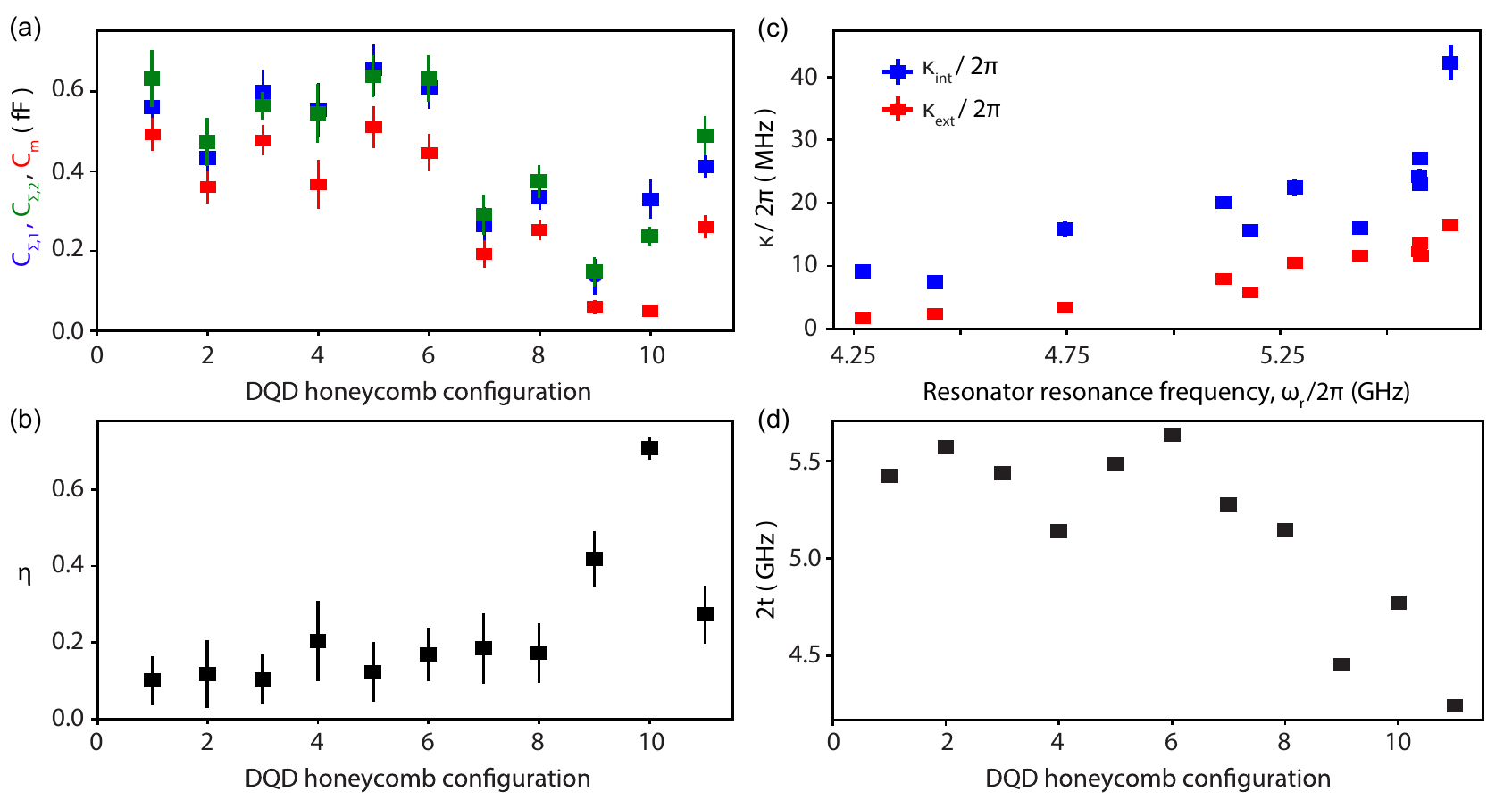}
	\caption{
	Comparison of some parameters of the eleven configurations analyzed in the main text.
	(a) DQD capacitances $\capTotal{1}$, $\capTotal{2}$ and $\capMut$.
	(b) Dipole strength $\eta$.
	(c) $\kappa_{\rm{ext}}$ and $\kappa_{\rm{int}}$, extracted by fitting the reflectance of the bare SQUID array to a Lorentzian with the DQD deeply in a Coulomb blockade.
	(d) Interdot tunneling rates $\Delta / h$ obtained from the JC model [see dashed lines in Fig.~\ref{fig3:Display_porperties}(a-c)].
	In (c) the data are ordered according to the resonator frequency. In remaining panels, the x axis is the configuration index.
	}
	\label{figKappas}
\end{figure*}

In the following, we describe how the dipole strength $\eta$ and the set of capacitance parameters
$\CL{1}, \CR{1}, \capTotal{1}, \CL{2}, \CR{2}, \capTotal{2}$ and $\capMut$ is determined from the DQD charge stability diagram. Here $\CL{i}$ [$\CR{i}$] is the capacitance between the left [right] side plunger gate and the $i$th dot and $\capTotal{i}$ is the total capacitance of the $i$th dot. $\capMut$ describes the inter dot capacitance. 
Together with the tunneling amplitude $\tunneling$, these parameters completely characterize the DQD system in our simplified model.
A summary of the extracted parameters for the eleven studied DQD configurations is given in Tab.~\ref{tab1:para_fig3_large}. Some of these parameters are also plotted in Fig.~\ref{figKappas}.
We could not measure the capacitances between the resonator gate and the QDs (the resonator gate lever-arm) since this gate is galvanically connected to ground via the resonator and thus cannot be DC-biased.

The voltages applied to the T and CG gates [see Fig.~\ref{fig1:SampleAndCircuit}(e)] are changed over hundreds of mV in order to realize the change of $C_m/C_{\Sigma}$ necessary to explore the different $\eta$ reported in this work, 
while typically smaller changes of a few mV are applied to fine tune the interdot tunneling rate $\tunneling/h$ by a few GHz, in order to realize the resonant condition with the resonator.

\begin{figure}[h]
	\centering
\includegraphics[width=\columnwidth]{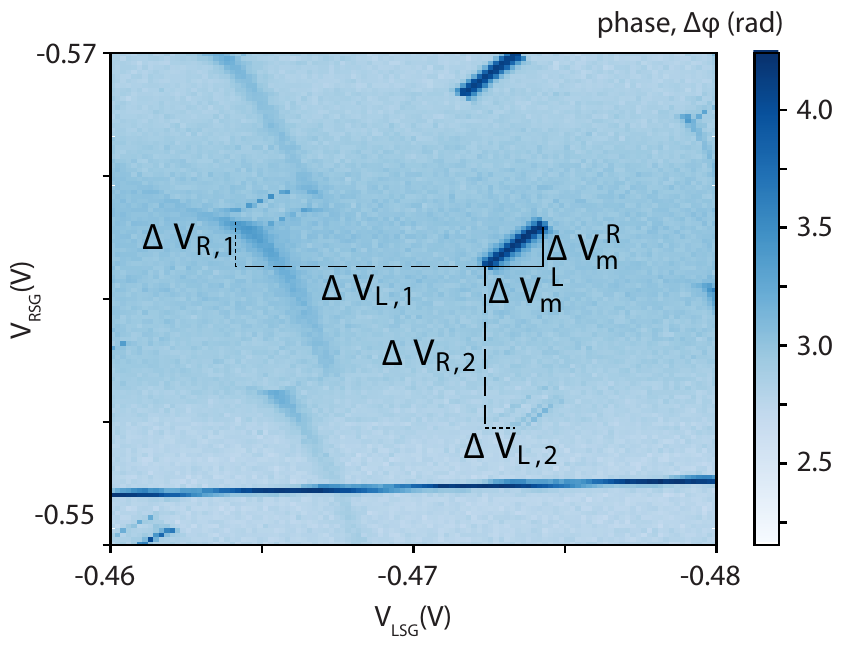}
	\caption{An example of a DQD charge stability diagram. It shows the phase response of the resonator reflectance while changing the voltage of gates RSG and LSG (see Fig.\ref{fig1:SampleAndCircuit}(e)). The six voltage differences indicated allow to extract the QDs capacitances and the dipole strength $\eta$.\label{fig:charge_stability}}
\label{fig:extract eta}
\end{figure}

Consider the dashed lines in the charge stability diagram~\cite{vanderwiel2002} in Figure~\ref{fig:charge_stability}. They represent the plunger gate voltage differences between two consecutive sets of triple points for which the only difference is that the effective charge of one dot changes by one electron charge $e$, while the total electro-static energy remains constant.
Hence one finds the four equations
\begin{widetext}
\begin{equation}
\begin{pmatrix}
\CL{1} \\
\CR{1}\\
\CL{2}\\
\CR{2}
\end{pmatrix}=
\begin{pmatrix}
\dVL{1} & -\dVR{1} & 0 & 0\\0 & 0 & \dVL{1} & -\dVR{1} \\ -\dVL{2}& \dVR{2} & 0 & 0\\ 0 & 0 & -\dVL{2} & \dVR{2}
\end{pmatrix}^{-1} \cdot \begin{pmatrix}
e \\
0\\
0\\
e
\end{pmatrix}
\end{equation}
\end{widetext}
where the voltage differences $\dVL{1}$, $\dVR{1}$, $\dVL{2}$ and $\dVR{2}$ are given by the length of the dashed lines in Figure~\ref{fig:charge_stability}.
The charging energy, which is given by Eq.~\eqref{eq:chargingenergy} in the main text, can be rewritten as
\begin{widetext}
\begin{equation}
E_{n_1,n_2}(\nGate{i})=\eC{1} (n_1 - \nGate{1})^2 + \eC{2}  (n_2-\nGate{2})^2 + \eC{\rm m}  (n_1-\nGate{1})(n_2-\nGate{2}),
\label{eq:EC}
\end{equation}
\end{widetext}
where $n_i$ is the number of electrons in dot $i$. Here, we introduced $\nGate{i}$ representing the effective number of electrons induced on dot $i$ by the voltages on the gates. In our experiment, a voltage change on the left (right) side gate, denoted by $\Delta V^L$ ($\Delta V^R$), results in a change $\Delta\nGate{1}$ ($\Delta \nGate{2}$) of $\nGate{1}$ ($\nGate{2}$) according to
\begin{equation}
\begin{pmatrix}
\Delta \nGate{1}\\
\Delta \nGate{2}
\end{pmatrix}= \frac{1}{e} \begin{pmatrix}
\CL{1} & \CR{1}\\
\CL{2} & \CR{2} 
\end{pmatrix}
\cdot 
\begin{pmatrix}
\Delta V^\mathrm{L}\\
\Delta V^\mathrm{R}
\end{pmatrix}. \label{eq:V_to_n}
\end{equation}
and the charging energy matrix is represented by
\begin{equation}
\begin{pmatrix}
\eC{1}  & \eC{\rm m}/2 \\
\eC{\rm m}/2 & \eC{2} 
\end{pmatrix} = \frac{e^2}{2} \begin{pmatrix}
\capTotal{1}& - \capMut \\
- \capMut& \capTotal{2}
\end{pmatrix}^{-1}.
\label{eq:charging_energy_matrix}
\end{equation}

Now we consider the solid black lines in Figure~\ref{fig:charge_stability} that connect adjacent triple points which are split due to the interdot mutual capacitance $\capMut$.
In the following, we use them to extract $\capMut$, $\capTotal{1}$ and $\capTotal{2}$. Without losing generality, we consider the triple point at the intersects of the \{(0,0), (0,1), (1,0)\} charge stability regions. The electrostatic energy at these triple point is given by
\begin{equation}
E_{0,0}(\nGate{i}^{(1)})=E_{0,1}(\nGate{i}^{(1)})=E_{1,0}(\nGate{i}^{(1)}). \label{eq:energies_TP1}
\end{equation} Similarly, the charging energy at the adjacent triple point, at the intersect of the \{(1,1), (0,1), (1,0)\} charge stability regions, is given by
\begin{equation}
E_{1,1}(\nGate{i}^{(2)})=E_{0,1}(\nGate{i}^{(2)})=E_{1,0}(\nGate{i}^{(2)}). \label{eq:energies_TP2}
\end{equation}
The voltage differences between these two triple points are denoted by $\dVLm$ and $\dVRm$ (lengths of solid black lines in Figure~\ref{fig:charge_stability}). Plugging these voltage differences into Eq.~\eqref{eq:V_to_n} as $\Delta V^\mathrm{L}=\dVLm$ and $\Delta V^\mathrm{R}=\dVRm$, we calculate the difference of the effective electron numbers induced by the gates, $\Delta \nGate{1}^{(\rm m)}$ and $\Delta \nGate{2}^{(\rm m)}$ between the two triple points. 
%While Eq.~\eqref{eq:V_to_n} is valid in general, here the superscript (m) denotes the belonging to the voltage differences $\dVLm$ and $\dVRm$.
In order to calculate the three parameters $\capMut$, $\capTotal{1}$ and $\capTotal{2}$, additionally to Eqs.~\eqref{eq:energies_TP1} and~\eqref{eq:energies_TP2} 
we consider the following relation which allows to calibrate the energy scale in the DQD stability diagram:
\begin{equation}
\hbar \detuning= E_{1,0}-E_{0,1} \label{eq:detuningPP}
\end{equation} at a specific set of gate voltages. We measured $\detuning$ by two-tone spectroscopy of the charge qubit at one specific gate voltage configuration and label the difference in the voltage on the left (right) gate between this configuration and the zero-detuning configuration by $\dVLd$ ($\dVRd$). By plugging these voltage differences into Eq.~\eqref{eq:V_to_n} as $V^\mathrm{L}=\dVLd$ and $V^\mathrm{R} =\dVRd$, we again convert the voltage differences into differences in the effective number of electrons induced by the gates which we call $\Delta \nGate{1}^{(\detuning)}$, $\Delta \nGate{2}^{(\detuning)}$. 
Here, the superscript ($\detuning$) highlights the correspondence to one specific set of $\detuning$, $\dVLd$ and $\dVRd$.

Combining Eqs.~\eqref{eq:EC},~\eqref{eq:energies_TP1},~\eqref{eq:energies_TP2} and~\eqref{eq:detuningPP}, we find the charging energies as
\begin{widetext}
\begin{equation}
\begin{pmatrix}
\eC{1}  \\
\eC{2}  \\
\eC{\rm m} 
\end{pmatrix} = \begin{pmatrix}
\Delta \nGate{1}^{(\rm m)} & 0 & \left(\Delta \nGate{2}^{(\rm m)}-1\right)/2 \\
0 & \Delta \nGate{2}^{(\rm m)} &  \left( \Delta  \nGate{1}^{(\rm m)}-1\right)/2 \\
-2 \Delta \nGate{1}^{(\detuning)} & 2 \Delta \nGate{2}^{(\delta)} &\Delta \nGate{1}^{(\detuning)} - \Delta \nGate{2}^{(\detuning)} 
\end{pmatrix}^{-1} \cdot \begin{pmatrix}
0 \\ 0 \\ \hbar \detuning
\end{pmatrix}.
\end{equation}
\end{widetext}
From the charging energies, the capacitances $\capTotal{1}$, $\capTotal{2}$ and $\capMut$ are then found using
Eq.~\eqref{eq:charging_energy_matrix}.
Finally, using Eq.~\eqref{eq:sensitivity} from the main text, we find the dipole strength as main result of this appendix, 
\begin{equation}
\eta=\frac{1-2\capMut/(\capTotal{1}+\capTotal{2})}{1+2\capMut/(\capTotal{1}+\capTotal{2})}, \label{eq:etacomplicated}
\end{equation}
where the capacitance parameters are given as
\begin{widetext}
\begin{align}
\capMut=&\frac{e^{2}\left(\dVLd \dVRm + \dVLm \dVRd\right)}{\hbar \detuning \left(\dVL{1} \dVR{2} - \dVL{2} \dVR{1}\right)}, \label{eq:capMut} \\
\capTotal{1}=& \frac{e^{2} \left(\dVLd \dVRm + \dVLm \dVRd\right) \left(\dVL{1} \dVR{2}-\dVLm \dVR{2} - \dVL{2} \dVRm - \dVL{2} \dVR{1}\right)}{\hbar \detuning \left(\dVLm \dVR{1} + \dVL{1} \dVRm\right) \left(\dVL{1} \dVR{2} - \dVL{2} \dVR{1}\right)}, \label{eq:capTot1}\\
\capTotal{2}=&\frac{e^{2} \left(\dVLd \dVRm + \dVLm \dVRd\right) \left(\dVL{1} \dVR{2}-\dVLm \dVR{1}- \dVL{1} \dVRm - \dVL{2} \dVR{1}\right)}{\hbar \detuning \left(\dVLm \dVR{2} + \dVL{2} \dVRm\right) \left(\dVL{1} \dVR{2} - \dVL{2} \dVR{1}\right)} .\label{eq:capTot2}
\end{align}
\end{widetext}
Note that when calculating $\capMut/\capTotal{i}$ 
%that is when dividing Eq.~\eqref{eq:capMut} by Eq.~\eqref{eq:capTot1} or by Eq.~\eqref{eq:capTot2}, 
the term  $\left(\dVLd \dVRm + \dVLm \dVRd\right)/\hbar \detuning$ cancels.
Hence,  $\capMut/\capTotal{i}$ and $\eta$ can be determined directly from the charge stability diagram without considering the energy calibration step.
%detuning, $\detuning$ for a specific voltage configuration.

In the simplified hypotethis of identical dots, $\capTotal{1}=\capTotal{2}=\capEqual$, with a symmetric coupling to their respective gates, $\CL{1}=\CR{2}$, and neglecting cross-gate capacitances, $\CR{1}=\CL{2}=0$, the expressions further simplify to
\begin{align}
\frac{C_\mathrm{m}}{C_\Sigma}&=\frac{\Delta V_\mathrm{m}}{\Delta V_\mathrm{g}-\Delta V_\mathrm{m}},\label{eq:CmdivDsigma}\\\eta&=1 - \frac{2\Delta V_{\rm m}}{\Delta V_{\rm g}},  \label{eq:eta_simple}
\end{align}
where $\Delta V_{\rm m}/\sqrt{2}\equiv\dVLm=\dVRm$ and $\Delta V_{\rm g}/ \sqrt{2}\equiv\dVL{1}=\dVR{2}$.

The error bars assigned to the extracted capacitances and of $\eta$ are determined by attributing, in the above procedure, an uncertainty to the positions of the four triple point in the stability diagram (see Fig.~\ref{fig:extract eta}). The errors were then propagated to the final results in Eqs.~\eqref{eq:etacomplicated}, \eqref{eq:capMut}, \eqref{eq:capTot1} and \eqref{eq:capTot2}.

\section{Considerations on the definition of the dipole strength $\eta$ in Eq.~\eqref{eq:suppression}.}
Here we report some considerations about the definition of the dipole strength for a DQD, introduced in Eq.~\eqref{eq:suppression}.
First of all, $\eta$ is dimensionless and independent on scales, such as the dot size or material constants.
Second, since $\capEqual \geq \capMut \geq 0$, its value\deleted{s} ranges between zero and one.
Third, we notice that zero mutal capacitance, $\capMut / \capEqual \to 0$, gives no suppression, $\eta \to 1$, and maximal mutual coupling $\capMut / \capEqual \to 1$ gives perfect suppression $\eta \to 0$.
Here it is useful to point out two possible limiting scenarios of increasing the interdot coupling to $C_m \to C_\Sigma$. Among other options, one can take this limit with either $C_\Sigma$ or $C_{\rm out}$ fixed.
In the former, the numerator in Eq.~\eqref{eq:suppression} is decreasing, reflecting the sum rule in Eq.~\eqref{eq:sumRule} as discussed in above. The numerator is constant in the latter, and its only role is to render the dipole strength dimensionless and normalized to one. In this case, one could omit the numerator from Eq.~\eqref{eq:suppression} to quantify the suppression effects. Nevertheless, keeping the numerator covers all possible scenarios together.
Finally, and what we deem most important, the definition of $\eta$ as given in Eq.~\eqref{eq:suppression} is practical:
the quantities defining $\eta$ can be directly read off the standard charging diagram of the double dot, as illustrated in Fig.~\ref{fig2:four_stability_diagramss} and Fig.~\ref{fig:extract eta}.

For illustration, we make the analogy with the useful microscopic model typically used to describe the origin of  
the coupling of the DQD electrical dipole moment $e\times d$ to the electrical field $\mathcal E$ generated by the resonator. 
In this case we can write the coupling term as
\begin{align}
	\qubitResonatorCoupling =\eta  \times   \qubitResonatorCoupling_{0} \equiv \eta \times e\times d \times \mathcal E \,,
	\label{eq:Dipole}
\end{align}
where 
we identified the \emph{bare} dipole energy of the DQD and resonator as $\qubitResonatorCoupling_{0} = e d\: \mathcal E = 2 e\: \sqrt{\frac{\hbar \resonatorAngularFrequency}{2 C_{\rm r}}} \frac{\capGate{1} - \capGate{2}}{C_{\rm out}}$, defined through bare quantities $e$, $d$ and $\E$. 
Equation~\eqref{eq:Dipole} thus expresses the coupling strength as the dipole energy arising from displacing an electron by distance $d$ in the electric field $\mathcal E$,
modified by the dipole strength $\eta \in \langle 0,1 \rangle$. 
Such a definition anticipates the three different possible micoscopic origins of  the dipole strength for the dipolar interaction: 
the dot background (core) electrons can partially screen the electric field acting on the hopping (valence) electron ($\eta \times \mathcal E $);
screening effects can reduce the effective hopping charge ($\eta \times e$); electrostatic tuning of the system may result in a configuration with reduced effective interdot distance ($\eta \times d$). 
%On the other hand, this definition is entirely impractical, since neither the hopping charge, $d$ nor $ \E$ is experimentally accessible and the three effect cannot be disentangled. 
Experimentally, we cannot distinguish these scenarios. We refer to them collectively as \emph{renormalization of the dipolar coupling energy}.
Equation~\eqref{eq:coupling} defines this dipole coupling $g$ using more accessible parameters.

\section{Detuning sensitivity to charge and voltage fluctuations}
\label{app:sensitivity}

In this appendix, we show how the DQD detuning energy responds to a change in the electrostatic environment, that is, if a voltage or a charge of an impurity in the dot environment changes. Our goal is to shed light on Eq.~\eqref{eq:sensitivity}, especially in the case where the two dots are not equal. The first line of Eq.~\eqref{eq:sensitivity} can be cast into 
\be 
\delta \detuning = e \delta V_\textrm{G} \frac{C_\textrm{G}\left[ d_\Sigma C_\Sigma + d_\textrm{G} (C_\Sigma -\capMut) \right]}{C_\Sigma^2(1-d_\Sigma^2/4)-\capMut^2}.
\label{eq:detuningChange} \ee
To arrive at this expression, we have introduced $C_\textrm{G} = (\capGate{1}+\capGate{2})/2$ and $C_\Sigma = (\capTotal{1}+\capTotal{2})/2$ for the average capacitances, and $d_\textrm{G} = (\capGate{1} - \capGate{2})/C_\textrm{G}$ and $d_\Sigma = (\capTotal{1} - \capTotal{2})/C_\Sigma$ for fractional differences. 
The formula further simplifies upon introducing ``polarizations'' of the dot capacitances to the gate and to the outside of the DQD, $\capOutDot{d} = \capTotal{d} - \capMut$. Namely, we define the polarizations
\be
P_\textrm{G} = \frac{\capGate{1}-\capGate{2}}{\capGate{1}+\capGate{2}}, \quad P_{\rm out} = \frac{\capOutDot{1}-\capOutDot{2}}{\capOutDot{2}+\capOutDot{2}}.
\ee
They relate to the fractional differences by $P_\textrm{G} = d_\textrm{G} / 2$ and $P_{\rm out} = d_\Sigma C_\Sigma/ 2(C_\Sigma - \capMut)$ and they take values between -1 and 1. The value $P_\textrm{G} \approx 1$ means that the magnitude of the left dot capacitance to the gate $V_\mathrm{G}$  is much larger than that of the right dot and analogously for $P_\mathrm{out}$. 
Since we aim at the leading order result, we neglect the $d_\Sigma^2/4$ term in the denominator of Eq.~\eqref{eq:detuningChange}, being higher-order in the difference of the two total capacitances. With that, and using the polarizations, the detuning change becomes
\be 
\delta \detuning = e \delta V_\textrm{G} \frac{\capGate{1}+\capGate{2}}{C_\Sigma+\capMut} \left( P_\textrm{G} + P_{\rm out} \right).
\label{eq:detuningChange2} \ee
This is the desired generalization of the second line of Eq.~\eqref{eq:sensitivity}: The difference of the two dots gives rise to an additional polarization, $P_\textrm{out}$. Using Eq.~\eqref{eq:detuningChange2} instead of Eq.~\eqref{eq:detuningChange}, the expression in Eq.~\eqref{eq:sensitivity2} would read
\begin{align}
	\delta \detuning = e V_G \, \frac{\capGate{1} + \capGate{2}}{\capOut} \left( P_\textrm{G} + P_{\rm out} \right) \eta\,,
\end{align}
where $C_\textrm{out}=C_\Sigma - C_\textrm{m}$ and the last term is the dipole strength as given in Eq.~\eqref{eq:suppression}. In other words, our definition of $\eta$ remains the same even if the dots are not equal.

We now derive the detuning change with respect to a charge impurity fluctuation. Concerning the electrostatic description, a charge impurity is an object similar to a dot: its primary variable is the charge and the voltage is a derived variable. Postponing the derivation and discussion of a model containing charge impurities to a separate publication, we state here only the result; the analog of Eq.~\eqref{eq:detuningChange} upon changing the impurity $i$ charge by $\delta Q_i$ is
\be 
\delta \detuning = e\frac{\delta Q_i}{\capTotal{i}} \frac{C_{i,1}+C_{i,2}}{C_\Sigma+\capMut} \left( P_i + P_{\rm out} \right),
\label{eq:detuningChange3} \ee
where $C_{i,d}$ is the capacitance between the impurity $i$ and the dot $d$, the polarization of these capacitances is $P_i = ( C_{i,1}-C_{i,2} )/({C_{i,1}+C_{i,2}})$, and $\capTotal{i}$ is the impurity self-capacitance. We conclude that there is a complete analogy between Eq.~\eqref{eq:detuningChange2} and Eq.~\eqref{eq:detuningChange3} upon interpreting $\delta Q_i /\capTotal{i}$ as the equivalent voltage fluctuation.

\section{How do figures of merit scale with the dipole strength \label{app:scaling}}
Here we discuss how the most important circuit-QED figures of merit scale with the dipole strength $\eta$. The scaling depends on whether the noise spectrum is singular (diverges at zero frequency) or regular and whether it couples to the qubit energy linearly or quadratically. Using the results of Ref.~\onlinecite{Ithier2005}, Eqs.~\eqref{eq:HDQD}, \eqref{eq:sensitivity2}, and \eqref{eq:suppression}, we obtained Tab.~\ref{tab:decayScaling} and Tab.~\ref{tab:FOMScaling}. In the former, one can see that in any scenario both decoherence and relaxation get bigger as $\eta$ grows. We find two possible power-laws, linear and quadratic. The decoherence rate is linear in $\eta$ if the noise is singular, such as a 1/f noise, and couples to the qubit linearly, that is, the qubit is not at a sweet spot. The linear scaling of the decoherence is observed in our experiment. In all other scenarios, the decoherence should be quadratic in $\eta$. The relaxation rate is always quadratic in $\eta$. In any case, aiming at maximal coherence calls for minimizing $\eta$. This minimization was the essence of the first experiment, described in Sec.~\ref{sec:Coherence}. Turning to the latter table, the first line gives the qubit-resonator coupling as proportional to $\eta$. Maximizing the coupling, as in aiming at the strong coupling regime, requires to maximize $\eta$. It was the core of the second experiment, outlined in Sec.~\ref{sec:USC}. 

We now comment on two additional figures of merit, the quality factor and cooperativity. They contain the coherence time, which we take as the inverse of the decoherence rate given in Tab.~\ref{tab:decayScaling}. This means we assume that the scaling of the decoherence and the relaxation is the same (quadratic), or that the relaxation can be made negligible if they differ. Under this assumption, the quality factor might benefit from decreasing $\eta$, while the cooperativity from increasing it. Whether the benefit is realized depends on the character of noise. 

Concluding, Tab.~\ref{tab:FOMScaling} uncovers a surprisingly large number of scenarios: The chosen figure of merit, the noise character, whether the qubit can be robustly kept at a sweet spot, and whether the relaxation is dominating the decoherence, all play a non-trivial role. Their combination decides whether maximization or minimization of $\eta$ is to be strived for.

%\begin{widetext}
\begin{table*}[ht]
\begin{tabular}{c|ccc|ccc}
\hline\hline
decay&&&decay&\multicolumn{3}{c}{dependence on qubit}\\
process&coupling & noise & type & configuration & sensitivity & suppression \\
\hline\hline
\multirow{4}{*}{\rotatebox[origin=c]{0}{pure dephasing}} 
&linear & singular & Gaussian & $\frac{\detuning}{\sqrt{\detuning^2 + \tunneling^2}}$&$\partial_V \detuning$ &$\eta$\\
&linear & regular & exponential & $\frac{\tunneling^2}{\detuning^2 + \tunneling^2}$&$(\partial_V \detuning)^2$ &$\eta^2$\\
&quadratic & low-freq. & algebraic & $\frac{\tunneling^2}{(\detuning^2 + \tunneling^2)^{3/2}}$&$(\partial_V \detuning)^2$&$\eta^2$ \\
&quadratic & high-freq. & exponential& $\frac{\tunneling^2}{(\detuning^2 + \tunneling^2)^{3/2}}$ &$(\partial_V \detuning)^2$&$\eta^2$\\
\hline
&&&&&\\
relaxation &linear & resonant & exponential & $\frac{\tunneling^2}{\detuning^2 + \tunneling^2}$& $(\partial_V \detuning)^2$&$\eta^2$ \\
&&&&&\\
%\hline\hline&&&&&\\
%driving & linear & resonant & Rabi oscillations & $\frac{\tunneling}{\sqrt{\detuning^2 + \tunneling^2}}$& $(\partial_V %\detuning)$&$\eta$ \\
%&&&&&\\
\hline \hline
\end{tabular}
\caption{Qualitative dependence of the qubit decay rates on the qubit properties, including the dipole strength $\eta$. The ``coupling'' denotes the power with which the noise variable $\delta V$ changes the qubit energy, for example, quadratic means $\delta \hbar \qubitAngularFrequency \propto \delta V^2$. ``Singular'' means that the noise diverges at zero frequency, for example, 1/f noise. For a quadratic coupling, the low-frequency and high-frequency parts of the noise spectrum are defined with respect to the inverse evolution time. ``Resonant'' means that only the noise at the qubit frequency is relevant. ''Decay type'' denotes the functional form of the decay envelope, such as the one in Eq.~\eqref{eq:decoherence} which happens to be Gaussian. Qubit ``configuration'' comprises the detuning and tunneling, ``sensitivity'' denotes the derivative of the detuning with respect to the fluctuating voltage, $\partial_V\detuning \equiv \partial\detuning / \partial V_\textrm{G}$, which can be obtained from Eq.~\eqref{eq:sensitivity}. Finally, ``suppression'' denotes the scaling with Eq.~\eqref{eq:suppression}. An example how to read this table: For a system dominated by a regular noise coupled linearly to the qubit, the qubit coherence decay is exponential, with a pure dephasing rate $\Gamma_\varphi \propto \frac{\tunneling^2}{\detuning^2 + \tunneling^2} (\partial_V \detuning)^2 \eta^2$.}
\label{tab:decayScaling}
\end{table*}

\begin{table}
\begin{tabular}{cc|c|c}
\hline\hline
&&\multicolumn{2}{c}{dominant noise}\\
figure of merit& formula & linear-singular & \qquad other \quad\phantom{x} \\
%&&\\
\hline\hline&&\multicolumn{2}{c}{}\\
coupling to cavity &$\qubitResonatorCoupling $& \multicolumn{2}{c}{$\eta^1$} \\
&&\multicolumn{2}{c}{}\\\hline&&\\
coherence time  &$T_2^* = 1/\Gamma_\varphi$ &$\eta^{-1}$ & $\eta^{-2}$ \\
&&&\\\hline&&\\
quality factor  &$Q = \qubitResonatorCoupling / \Gamma_\varphi$ &$\eta^{0}$ & $\eta^{-1}$ \\
&&&\\\hline&&\\
cooperativity & $\qubitResonatorCoupling^2/\Gamma_\varphi \kappa$ & $\eta^{1}$ & $\eta^{0}$ \\
&&&\\\hline
\end{tabular}
\caption{Scaling of several figures of merit with the dipole strength. We used $g \propto \eta$ for the interaction strength, as follows from Eq.~\eqref{eq:coupling}, and the results from Tab.~\ref{tab:decayScaling} for the pure dephasing rate $\Gamma_\varphi$. Here we work in the hyphothesis that the decoherence is dominated by dephasing process ($\Gamma_2 \sim \Gamma_{\phi}$) which is a resonable assumption for a DQD charge qubits.}
\label{tab:FOMScaling}
\end{table}

\section{SQUID and Junction Array High Impedance Resonators} \label{sec:resonators}
High impedance resonators represent a valuable tool to increase the vacuum voltage fluctuations to maximize the coupling strength with the two-level electrical dipole moment. It allowed achieving the strong coupling regime for electrons confined in semiconductor DQDs~\cite{Stockklauser2017}.
For superconducting artificial atoms electrically coupled to the microwave radiation, it has been recently demonstrated that high impedance resonators enable to reach a much higher coupling strength that brings the system in the ultrastrong and deep strong coupling regimes \cite{Kockum2018a, FornDiaz2019}.

\begin{figure*}[!t]
	\includegraphics[width=\textwidth]{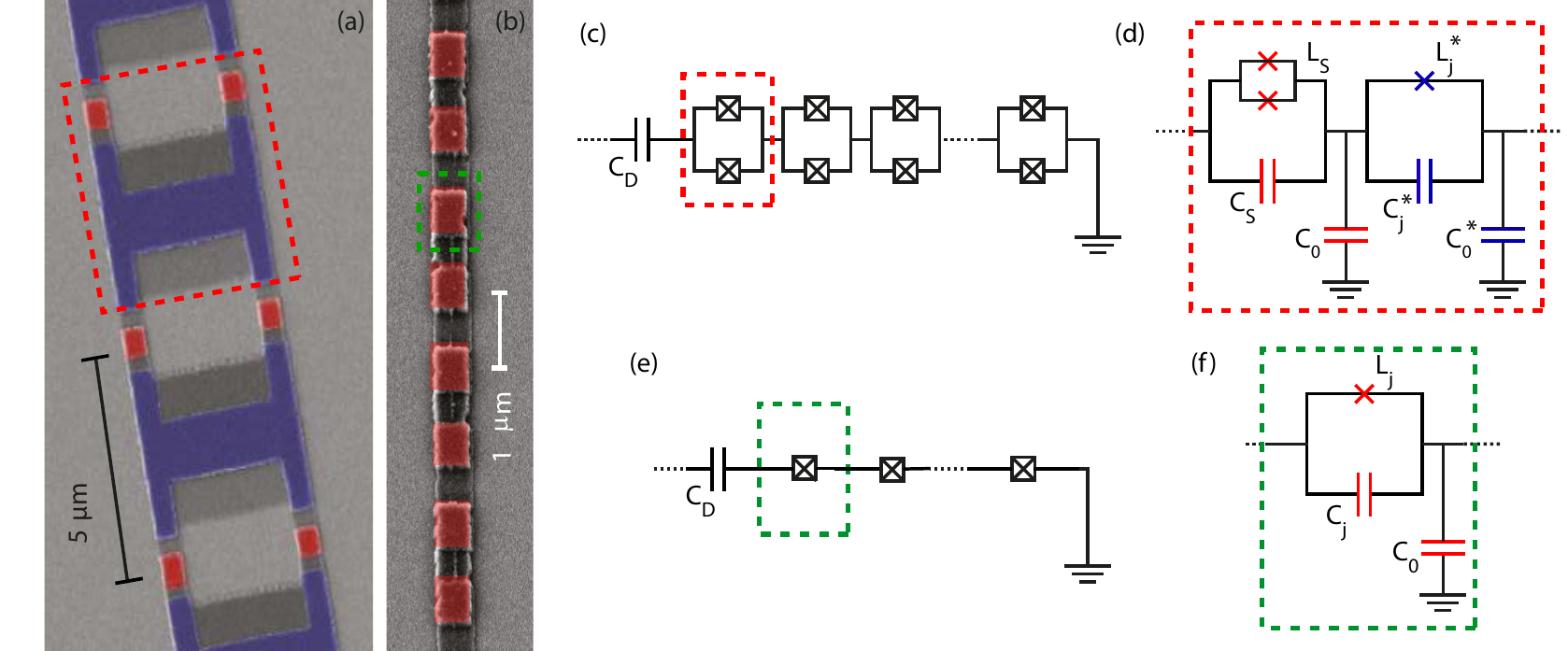}
	\caption{(a) False colored SEM micrograph of a section of a SQUID array. The dashed red line encloses the unit cell of the SQUID array.
	(b) False colored SEM micrograph of a section of a Josephson junction array. The dashed green line encloses the unit cell of the array, with a single $0.5 \times 0.9$ $\mu m^2$ Josephson junction.
	(c) Schematic circuit for a $\lambda/4$ SQUID array resonator. $C_{D} = C_{c} + C_{RPG} C_{g} $ represents the capacitive coupling between the resonator array and the microwave feedline, the DQD device, and the rest of the DQD gates. The other end of the array is grounded.
	(d) Circuit schematic of the unit cell of the SQUID array. $L_{S}$ and $L^{\star}_{J}$ represent the inductance of each SQUID junctions (red) and of the extra Josephson junction (blue) in series, 
	while $C_{S}$ (red) and $C^{\star}_{J}$ (blue) represent their junction capacitance. $C_{0}$ and $C^{\star}_{0}$ are their 	respective capacitance to the ground.	
	(e) Schematic circuit for a $\lambda/4$ JJ array.
	(f) Circuit schematic of the unit cell of the array. $L_{J}$ represents the Josephson inductance while $C_{J}$ and $C_{0}$ the junction capacitance and stray capacitance to ground, respectively.
	%For a josephson array the unit cell would be given by a parallel of $C_{J}$ and $L_{J}$
	}
	\label{Fig:Squid}
\end{figure*}

The SQUID and JJ array resonators, represented in Fig.~\ref{Fig:Squid}, are 1D Josephson junction metamaterials with a multimode spectrum \cite{Masluk2012}. Their design parameters have to be chosen appropriately to exhibit the resonance frequency of the array fundamental mode lying within the measurement bandwidth and well separated in frequency from the second mode of the array \cite{Masluk2012}.
In Fig.~\ref{Fig:Squid} (a) and (b) [(c) and (e)] we report a micrograph of [a circuit model for] the SQUID and JJ array resonators, respectively.
The base unit of the SQUID [JJ] array resonator is enclosed by the dashed red [blue] line in Fig.~\ref{Fig:Squid}(a), (c-d) [(b), (e-f)].
The fabrication process of the SQUID array, based on the shadow evaporation technique, generates the two small Josephson junctions in parallel (the SQUID junctions, in red) that are in series with an extra larger junction (in blue), with $\sim 11$ times more extended area, as we can see in Fig.~\ref{Fig:Squid}(a).

We realized SQUID junctions with inductance $L_{\rm{S}} \sim 1.25 ~\rm{nH}$ and capacitance $C_{\rm{S}} \sim 80 ~\rm{fF}$, while the large junctions have $L_{\rm{J}}^{\star} \sim 0.11 ~\rm{nH}$ and $C_{\rm{J}}^{\star} \sim 880 ~ \rm{fF}$.
Each section of the SQUID array presents on average a stray capacitance to ground of $C_0\sim C_{\rm{gnd}}/N=C_0^{\rm{J}}+C_0^{\rm{S}}$ (see table~\ref{table:nonlin}), where $C_0^{\rm{J}} \sim 6 C_0^{S}$ is the average capacitance to ground of the series junction.
Therefore, the part of the base unit containing this extra junction dominates the stray capacitance to the ground per section but adds a negligible contribution to the total array inductance. This limits the impedance reachable by the fundamental mode of the resonator array.

We can model these arrays as distributed $\lambda/4$ resonators, being shunted to ground on one end [see Fig.~\ref{fig1:SampleAndCircuit}(c) and (f)]. The capacitance between the array resonator and the right QD, the microwave feedline and the rest of the DQD depletion gates are estimated to be $C_{\rm{RPG}} \sim 0.07 \, \rm{fF}$, $C_{\rm{c}} \sim 3 \, \rm{fF}$ and $C_{\rm{g}} \sim 1.5 \, \rm{fF}$, respectively.
% [see Fig.~\ref{Fig:Squid}(c,e)].
%All of the capacitances reported above are estimated using numerical simulations with Maxwell.

As represented in Fig.~\ref{Fig:Squid}(f) and reported in Table~\ref{table:nonlin}, we can model each unit cell of the JJ array with a parallel of an inductance $L_{\rm{J}} \sim 1.5 \, \rm{nH}$ and capacitance $C_{\rm{J}}\sim 40  \, \rm{fF}$, in series to a capacitance $C_0^{\rm{JJ}}$ to ground. 
For $N=70$ junctions in series we obtained a total array resonator length of about $70 \, \mu \rm{m}$, with an estimated total array inductance of $L_{\rm{tot}} \sim 102 \, \rm{nH}$ and a total stray capacitance to ground of $C_{\rm{gnd}} \sim 4.9 \, \rm{fF}$.
This allows to estimante a JJ array resonator impedance $Z_{\rm{r}} \sim \sqrt{L_{\rm{tot}} / (C_{\rm{gnd}} + C_{\rm{c}} + C_{\rm{g}}+C_{\rm{RPG}} )} \sim 3.8 \, \rm{k}\Omega$, almost four times higher than the SQUID array impedance, which allows to increase the coupling strength with the DQD electric dipole moment of a factor $\sqrt{ Z_{\rm{r}}^{\rm{JJ}} / Z_{\rm{r}}^{\rm{Sq}} } \sim 2$.

\begin{table}[ht]
\centering % used for centering table
\begin{tabular}{|c| c| c|} % centered columns (4 columns)
\hline %inserts double horizontal lines
 & SQUID Array & Junction Array \\ [0.5ex] % inserts table
%heading
\hline\hline % inserts single horizontal line
$Z_{\rm{r}}$ ($\rm{k}\Omega$) & 1.1 & 3.8 \\ \hline% inserting body of the table
$\omega_{\rm{r}}/2\pi$ (GHz) & 6.2 (tunable) & 5.665 \\ \hline
$\kappa_{\rm{int}}/2\pi$ (MHz) & Fig.\ref{figKappas}(c) & 23.0 \\ \hline
$\kappa_{\rm{ext}}/2\pi$ (MHz) &  Fig.\ref{figKappas}(c)  & 4.0 \\ \hline
$N$ & 34 & 72 \\ \hline
$\omega_{\rm{p}}/2\pi$ (GHz) & 16.6 & 16.1 \\ \hline
Length ($\mu \rm{m}$) & 200 & 70 \\ \hline
$K_{00}$ (kHz) & 5 & 60 \\ \hline
$L_{\rm{tot}}$ (nH) & 31 & 102 \\ \hline
$C_{\rm{gnd}}$ (fF) & 19 & 5 \\ \hline
$C_{\rm{c}}$ (fF) & 2.5 & 1.5 \\ \hline
$C_{\rm{g}}$ (fF) & 1.5 & 1.5 \\ [1ex] % [1ex] adds vertical space
\hline %inserts single line
\end{tabular}
\caption{Comparison between SQUID and JJ array resonators} % title of Table
\label{table:nonlin} % is used to refer this table in the text
\end{table}

\section{Master equation - dissipative dynamics of DQD and resonator}\label{sec:masterEq}
%\pss{please check if some those initial equations have been already defined before and maybe cite those? Then, I am not sure about the factor 4 in $\omega_{q} = \sqrt{\detuning^{2} + 4 \tunneling^{2}}$ compare to Eq.1}
Here, we give a short introduction into the theoretical modeling of the experimental data directly. The model includes the double quantum dot, the resonator, and the microwave drive through a transition line.
For simplicity of notation, we use the convention $\hbar = 1$ in this section.
The dissipative dynamics of the system is described by the master equation
\begin{align}
	\dot \rho = -i \comm{H}{\rho} +\sum_{k} \mathcal L_{k} \rho \label{eq:METotal}\,,
\end{align}
where $H$ is the Hamiltonian of the system and the $\mathcal L_{k} \rho$ describe different dissipative channels introduced in the following. 
\subsection{Hamiltonian}
%\pss{please check if the first part and equations are needed or are repetition of something already defined before that could be cited} %The full system is described DQD qubit and the $\lambda/4$ resonator. 
The double quantum dot is well described by the Hamiltonian
\begin{align}
	H_{\text{DQD}} = \frac12 \detuning \sigma_{z} + \frac12 \tunneling \sigma_{x} = \frac12 \omega_{q} \tilde\sigma_{z}\,,
\end{align}
with the Pauli matrices $\sigma$ in the DQD position basis and $\tilde\sigma$ in its eigenbasis, and where $\detuning$ is the detuning and $\tunneling$ is the tunnel splitting between the two dots.
The DQD level splitting is $\omega_{q} = \sqrt{\detuning^{2} + \tunneling^{2}}$.
The resonator is described by
\begin{align}
	H_{\text{res}} = \omega_{\text r} a\hc a \,,
\end{align}
with its resonance frequency $\omega_{\text r}$ and the bosonic annihilation operator $a$. 
The coupling between DQD and resonator is between the quantum dots dipole moment and the electric field of the harmonic oscillator mode, so we write 
\begin{align}
	H_{\text{DQD-res}} = g_{0} \sigma_{z} (a + a\hc) = g_{0} \left( \cos{\varphi} \tilde\sigma_{z} + \sin{\varphi} \tilde\sigma_{x} \right) (a + a\hc) \,,
\end{align}
with the DQD mixing angle $\tan{\varphi} = \frac{\detuning}{\tunneling}$.
The total system Hamiltonian is then
\begin{align}
	H= H_{\text{DQD}} + H_{\text{res}} + H_{\text{DQD-res}} \label{eq:HTotal}\,.
\end{align}

\subsection{Dissipative processes}
The quantum dot and resonator are unavoidably coupled to the environment, leading to energy loss and dephasing. 
For the resonator, incoherent photon loss can be described in the master equation through a dissipative term 
\begin{align}
	\mathcal L_{\text{res}} \rho = \kappa_{\text{int}} \diss{a}\rho \,,
\end{align}
with the internal photon loss rate $\kappa_{\text{int}}$.
In practise, the resonator decay is made up of an internal component $\kappa_{int}$, stemming from coupling to the intrinsic environment, and an external coupling rate, $\kappa_{ext}$, stemming from coupling to external modes, such as the transmission lines used for driving.
In our treatment here, the external coupling will be taken into account through the SLH cascading of an external driving field, described in the next section, so that here we only include the intrinsic losses $\kappa_{int}$.
For the DQD, we assume a transversal decay channel, leading to energy relaxation at rate $\Gamma_{1}$, as well as a pure dephasing process due to fluctuations in the level splitting, leading to dephasing at rate $\Gamma_{2} = \frac12 \Gamma_{1} + \Gamma_{\varphi}$. 
The contributions to the master equation due to the dissipative dynamics of the DQD are then
\begin{align}
	\mathcal L_{\text{DQD}} \rho = \Gamma_{1} \diss{\tilde\sigma_{-}}\rho + \frac12 \Gamma_{\varphi} \diss{\tilde\sigma_{z}}\rho \,.
\end{align}

\subsection{SLH model - driven, dissipative dynamics of DQD and resonator}
We use the SLH cascaded quantum systems approach to model scattering of microwave photons in the transmission line off the $\lambda/4$ resonator~\cite{Combes:Adv:2017, Muller:PRA:2017, vanWoerkom2018}.
To this end, we cascade in a drive field for the resonator, which adds an effective drive term to the Hamiltonian as
\begin{align}
	H_{\text{drive}} = \frac{1}{2\ii} \sqrt{\kappa_{\text{ext}}} \left( \alpha\: a\hc - \alpha\cc\: a \right) \,,
	\label{eq:MESLHDrive}
\end{align}
where we assumed a single-sided, $\lambda/4$-type cavity driven with a coherent state of amplitude $\alpha$. 
Here we have additionally transformed the system into the rotating frame at the drive frequency $\omega_{\text d}$ of the coherent field input $\alpha$. 
The cascading also adds another dissipative part to the master equation, which describes the decay of the resonator modes into the transmission line, which is assumed to have a constant spectrum. 
This term is written as
\begin{align}
	\mathcal L_{\text{SLH}} \rho = \diss{\hat L} \rho 
	\label{eq:MESLH}
\end{align}
with the decay operator
\begin{align}
	\hat L = \sqrt{\kappa_{\text{ext}}} \: a + \alpha \mathds{1} \,.
\end{align}
Using this formalism, we can now calculate the amplitude $\beta$ and photon flux $n$ of the field scattered off the resonator as
\begin{align}
	\beta = \Tr{\left\{L \rho\right\}} \quad , \quad n = \Tr{\left\{L\hc L \rho\right\}}
\end{align}
where $\rho$ is the solution of the total master equation, Eq.~\eqref{eq:METotal}, now also including the drive and decay term from the cascading procedure, Eqs.~\eqref{eq:MESLHDrive} and ~\eqref{eq:MESLH}.
As equilibration of the field in the transmission lines happens typically very fast,  we can assume that scattering in experiments happens in the steady-state of the system, so that we only need to calculate the steady-state density matrix $\bar\rho$ for all cases. 

\subsection{Visibility of vacuum Rabi splitting}\label{sec:visibility}

\begin{figure*}[!htb]
	\includegraphics[width=\textwidth]{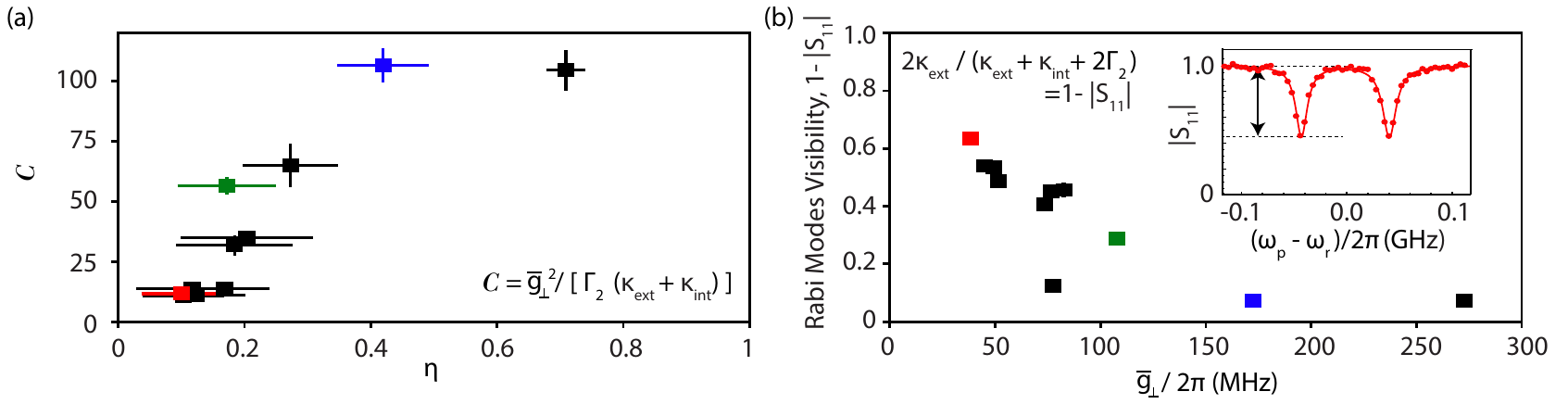}
	\caption{
	Extracted figures of merit of light-matter hybridization.
	(a) System cooperativity $C=\normCoupling^2/(\Gamma_2 \kappa)$. 
	(b) Visibility of the vacuum Rabi modes at resonance $(1-\lvert S_{11}\rvert )= 2\kappa_\mathrm{ext}/(\kappa_\mathrm{ext}+\kappa_\mathrm{int}+2\Gamma_2)$ \emph{vs.} the DQD-SQUID array coupling strength $\normCoupling$.
	}
	\label{figS2}
\end{figure*}

To find analytical expressions for the scattered field in the special case where DQD and resonator are tuned to resonance,
%and we observe the vacuum-Rabi mode splitting of the resonator, 
we take the analogy to the case of a two-level system embedded in a waveguide, c.f the supplementary material of Ref.~\cite{Hamann:PRL:2018}. 
For exact resonance between DQD and resonator, $\omega_{r} = \omega_{q} = \omega_{0}$, the eigenstates of the coupled system are $\ket\pm = \frac{1}{\sqrt{2}} \left( \ket{0,e} \pm \ket{1,g} \right)$. 
To make analytical progress, we are focussing on driving the transition between the total system groundstate $\ket{0,g}$ and one of the coupled eigenstates $\ket \pm$, 
analogous to the two-level system case. 
To this end we diagonalise the total Hamiltonian of the resonator plus DQD, and then consider the relevant operators 
in the diagonal basis, when reduced to a subset of states, i.e. the total system groundstate $\ket{0,g}$ and either of the two maximally mixed eigenstates $\ket{\pm}$.
For each of these transitions we write the input-output relations in the SLH formalism in analogy to the case of a driven two-level system, to find the reflectance of the $\lambda/4$-type resonator in resonance with the DQD.
For small drive amplitude $\alpha$ far from saturation, we find to lowest order in $\alpha$
\begin{align}
%	r_{\pm} = \beta / \alpha = 1 - \frac{2 \kappa_{\text{ext}}}{(\kappa_{\text{ext}} + \kappa_{\text{int}} + \gamma_{1} + 2 \gamma_{\varphi} + 4 i ( \omega_{0} - \omega_{\text d} \pm \frac12 g_{0}))} \,,
	r_{\pm} = \beta / \alpha = 1 - \frac{2 \kappa_{\text{ext}}}{(\kappa_{\text{ext}} + \kappa_{\text{int}} + 2 \Gamma_{2} + 4 i ( \omega_{0} - \omega_{\text d} \pm \frac12 g_{0}))} \,,
\end{align}
%\begin{align}
%	r_{\lambda/4} = 1 - \frac{2 \kappa_{\text{ext}}}{(\kappa_{\text{ext}} + \kappa_{\text{int}} + \gamma_{1} + 2 \gamma_{\varphi} + 4 i ( \omega_{0} - \omega_{d} - \frac12 g_{0}))}
%\end{align}
where $\omega_{\text d}$ is the frequency of the drive field and $g_{0}$ is the coupling strength between resonator and DQD.
%\pss{in the above equation we use $\gamma_{1}$ and $\gamma_{\varphi}$ but in the main paper we use $\Gamma_2$...could you please make the notation uniform?}
As we assume perfect resonance between DQD and resonator, the two expressions differ only in the position of the resonance. 
For resonant driving of either transition, i.e when $\omega_{d} = \omega_{0} \pm \frac12 g_{0}$, these reduce to 
\begin{align}
%	r_{\pm, \text{res}} &= 1-|S_{11}|= 1 - \frac{2 \kappa_{\text{ext}}}{(\kappa_{\text{ext}} + \kappa_{\text{int}} + \gamma_{1} + 2 \gamma_{\varphi})}  = 1 - \frac{\kappa_{\text{ext}}}{2 \gamma_{2, \pm}}\,.
	r_{\pm, \text{res}} = 1-|S_{11}|= &1 - \frac{2 \kappa_{\text{ext}}}{(\kappa_{\text{ext}} + \kappa_{\text{int}} + \Gamma_{1} + 2 \Gamma_{\varphi})}  = \nonumber \\ 
=&1 - \frac{\kappa_{\text{ext}}}{2 \Gamma_{2, \pm}}\,.
%	t_{\lambda/2, \text{res}} &= 1 - \frac{\kappa_{\text{ext}}}{(\kappa_{\text{ext}} + \kappa_{\text{int}} + \gamma_{1} + 2 \gamma_{\varphi})} 
\label{eq:visibility}
\end{align}
Thus, the depth of the reflection peak on resonance is given by the ratio of the external coupling of the resonator to twice the total linewidth of the DQD-resonator hybridised states, 
$\Gamma_{2, \pm} = \frac14 \left( \kappa_{\text{ext}} + \kappa_{\text{int}} + \Gamma_{1} + 2 \Gamma_{\varphi}\right)$, analogous to the case of scattering off a two-level system~\cite{Hamann:PRL:2018}.
A plot of the visibility of the Rabi modes, extracted according to Eq.~\ref{eq:visibility}, is reported as a function of $\eta$ in Fig.~\ref{fig3:Plot_parameters}(f) in the main text and as a function of the renormalized coupling strength $\normCoupling$ in Fig.~\ref{figS2}(b).

\subsection{Fits}
Peaks from experiments are fitted to the Hamiltonian level structure, i.e. the position of levels in Eq~\eqref{eq:HTotal}. 
When fitting the full transmission curve as function of frequency, the SLH model is used, where for simplicity we set $\gamma_{1} = 0$, as only the total DQD linewidth is relevant for these fits.

\section{System Cooperativity \label{Figures of Merit}}

A typical figure of merit for a cQED light-matter platform is represented by the cooperativity, defined as $C = \normCoupling^2/[\Gamma_2 (\kappa_{\rm{ext}}+\kappa_{\rm{int}})]$.
Introducing the cooperativity allows characterizing the
strength of light-matter interaction in our hybrid system and to compare it with what already achieved in previous experiments with similar hybrid devices \cite{Cottet2017b}. 
The strong coupling regime is represented by having a Cooperativity greater than unity.
In this case, the coupling is strong in the sense that at
resonance, nearly every photon entering the cavity is coherently transferred into the matter system.

In Fig.~\ref{figS2}(a), we reported the system cooperativity extracted for the eleven studied DQD configurations as a function of the dipole strength $\eta$.
We can notice how, despite increasing $\eta$ makes the DQD decoherence rate $\Gamma_2$ higher (see Fig.~\ref{fig3:Plot_parameters}), the cooperativity overall increases with $\eta$ too. This is in line with what is illustrated in the main part of the manuscript, where we reported that $\normCoupling,\,\Gamma_2\, \propto \eta$, therefore $C\propto \eta$.
In this work, making use of the described tuning strategy for the DQD electric dipole strength, we push the limits for the cooperativity achieved for the semiconductor QD-resonator hybrid device above 100, representing highest cooperativity measured so far for a QD-resonator hybrid system.
Furthermore, by adequately filtering the DQD gate lines has been shown that it is possible to keep a resonator linewidth $<1\, \rm{MHz}$ \cite{Mi2017c,HarveyCollard2020}, which, if implemented in our device, could allow achieving cooperativity up to $C \sim 1500$.

\section{Renormalization of the coupling strengths (Eq.~\ref{eq:normalization})}\label{secrenormalization}
In the following, we describe the strategy that we used to renormalize the coupling strengths extracted from the eleved studied DQD configurations in order to compare them (see Eq.~\ref{eq:normalization}).
The necessity to renormalize them comes from the fact that the hybridized spectra for the investigated DQD configurations have been taken not at exactly the same resonator frequency and DQD tunneling amplitude (see Table\ref{tab1:para_fig3_large}).

\begin{figure*}[!htb]
\includegraphics[width=\textwidth]{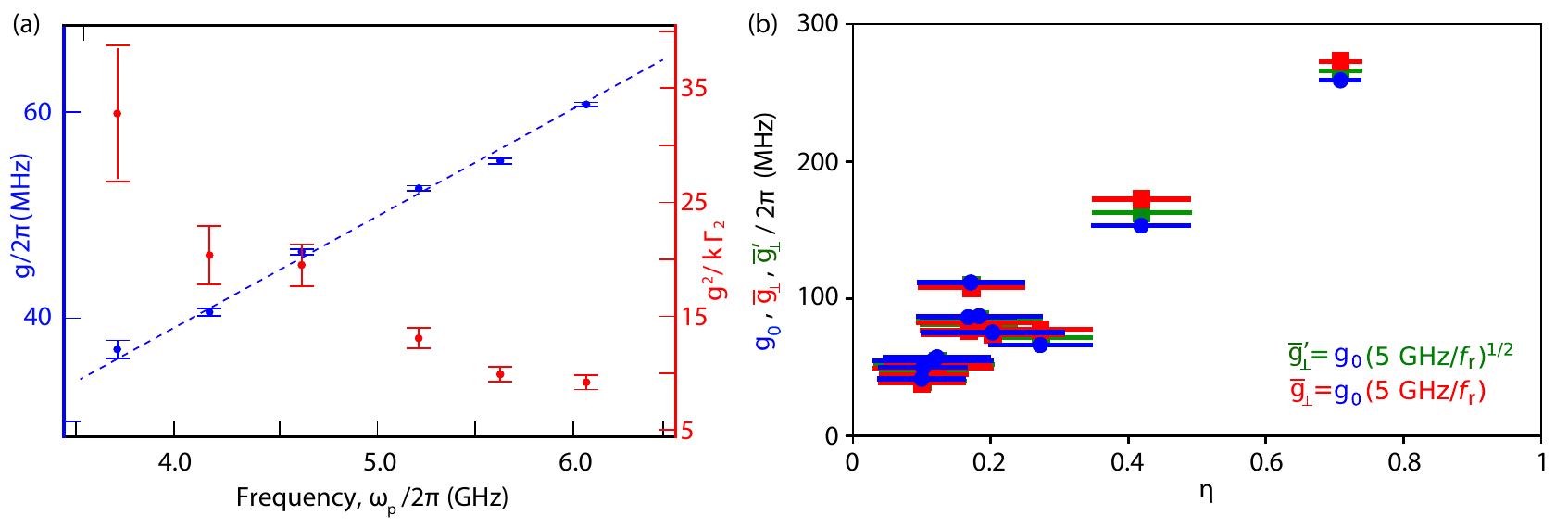}
	\caption{(a)
	 (Left axis) $g/2\pi$ extracted by measuring a Rabi mode splitting for the DQD qubit in resonance at $\detuning=0$ with the SQUID array fundamental mode, for different resonator frequency. (Rigth axis) system cooperativity, $g^2/(\kappa \Gamma_2)$, at different resonator frequency. During these measurements the DQD system is kept at the sweetspot $\detuning=0$.
(b) Comparison of the exctracted coupling strenghts corrected just for the mixing angle $g_0=g \Delta/\omega_r$ with the normalized $\normCoupling \propto g_0 \frac{5\, \rm{GHz}}{\resonatorAngularFrequency /2\pi}$ and $\normCoupling^{'} \propto g_0 \sqrt{ \frac{5\, \rm{GHz}}{\resonatorAngularFrequency /2\pi}}$.
	}
	\label{fig_glinear}
\end{figure*}

The first term in Eq.~\ref{eq:normalization}, $\tunneling/\omega_{\rm{r}}$, comes from the mixing angle renormalization of the DQD dipole strenght\cite{Frey2012} (see Eq.~\ref{eq:HDQD}).
In Fig.\ref{fig_glinear}(a) we report a study, coming from a similar device with a nominally identical DQD coupled to a SQUID array resonator, of the coupling strength between a DQD and the resonator as a function of the resonator frequency. It is realized by investigating the DQD and resonator hybridization (resonant condition $\omega_{\rm{r}} \sim \omega_{\rm{q}}$) by keeping the DQD at the sweetspot $\detuning=0$. The resonance frequency of the DQD is changed systematically by changing its interdot tunneling amplitude $\tunneling$ via the voltages applied to the depletion gates, and the frequency tunability of the SQUID array allows it to get in resonance with the DQD.

%. During this measurement, the DQD system is kept at the sweetspot $\detuning=0$.
The extracted evolution of the coupling rate $g$ as a function of the resonator frequency $\omega_{\rm{r}} \sim \omega_{\rm{q}}$ can be modeled accurately by a simple linear dependence (see blue dotted line in Fig.\ref{fig_glinear}(a)). 
Instead, considering that $Z_\mathrm{r}=1/(\omega_\mathrm{r} C_\mathrm{r})$, from Eq.~\ref{eq:coupling}
a $g\propto\sqrt{Z_\mathrm{r}}\omega_\mathrm{r} \propto \sqrt{\omega_\mathrm{r}}$ is expected assuming a simple lumped-element equivalent model of the resonator and that the tuning process of the inter-dot tunneling rate does not appreciably modify the DQD electrical dipole moment and its capacitive coupling to the resonator gate (QD level-arm).
The observed linear scaling of $g$ \emph{v.s.}\,$\omega_{\rm{r}}$ could suggest that other mechanisms take place in either the resonator impedance or the DQD electric dipole moment during the tuning procedures of the interdot tunneling and SQUID array resonance. The change in tunnel rate or DQD shape could present a considerable influence in the magnitude of the electrical dipole moment of the DQD, therefore of the coupling rate. A complete understanding of these mechanisms will require further investigations.

Fig.\ref{fig_glinear}(b) shows a comparison of the exctracted coupling strenghts corrected just for the mixing angle $g_0=g \Delta/\omega_r$ with the normalized $\normCoupling \propto g_0 (5\, \rm{GHz}/\resonatorAngularFrequency)$ and $\normCoupling^{'} \propto g_0 \sqrt{ 5\, \rm{GHz}/ \resonatorAngularFrequency}$. 
We notice how, for the dataset repored in this manuscript, the correction coming from the normalization choice does not exceed $10\%$ of the bare extracted couling rates.

\section{Extra data \label{Extra data}}

Here we report some extra measurements and dataset which the reader may find useful to better interpret the measurements reported in the main text.

\begin{figure*}[!htb]
	\includegraphics[width=\textwidth]{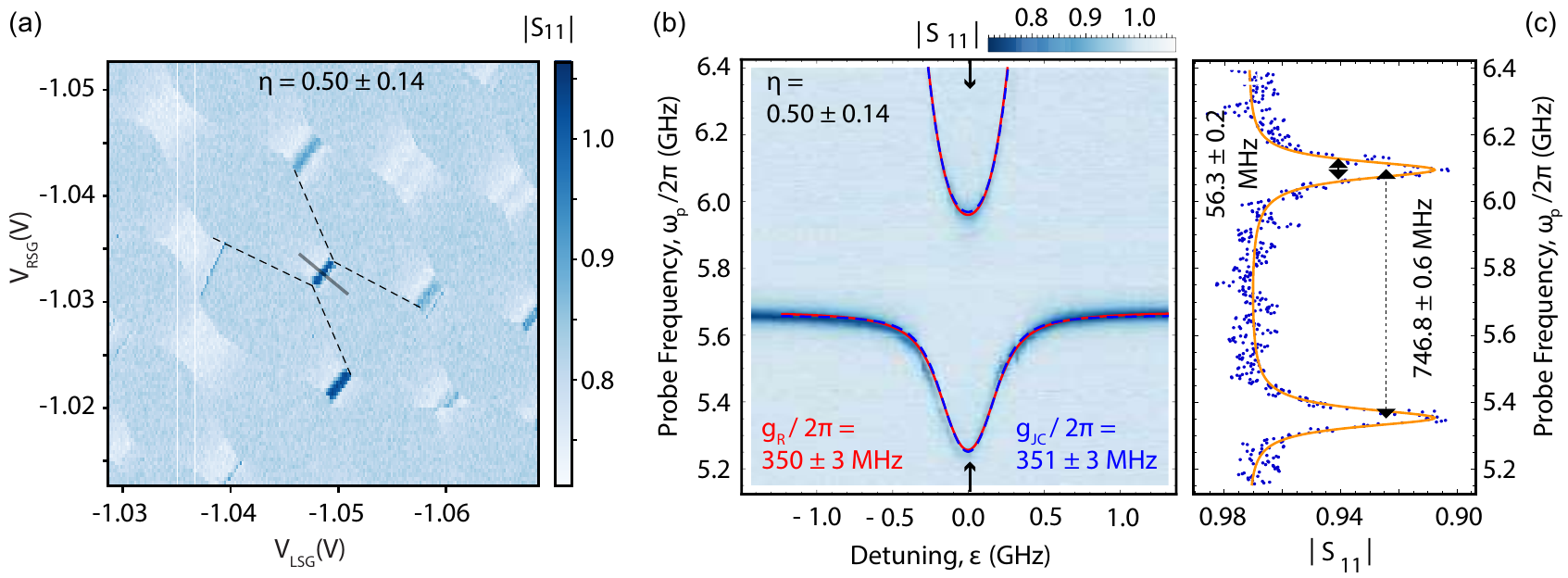}
	\caption{
		Investigation of a second DQD configuration with $\eta \sim 0.50 \pm 0.14$ for hybrid  DQD-Josephson junction array device represented in Fig.\ref{fig1:SampleAndCircuit}(f). 
		(a) Charge stability diagram of the DQD measured by monitoring the change in resonator reflectance amplitude $|S_{11}|$ for the extracted dipole strength $\eta \sim 0.50 \pm 0.14$.
		(b)Resonator amplitude response $|S_{11}|$ taken by varying the DQD detuning $\detuning$ along the grey line indicated in panle (a) by applying appropriately chosen voltages to the two side gates.
		Red (blue) line are independent fits to the Rabi (JC) model (see Appendix \ref{sec:masterEq}).
% from which we extract 
%		$\tunneling_{\rm{R}}/2\pi = 5.547 \pm 0.006$ GHz ($\tunneling_{\rm{JC}}/2\pi= 5.561 \pm 0.006$ GHz),
%		$\omega_{r,\rm{R}}/2\pi \sim 5.6580 \pm 0.0005~\rm{GHz}$ ($\omega_{r,\rm{JC}}/2\pi \sim 5.6594 \pm 0.0005~\rm{GHz}$)  
%		and $g_{\rm{R}}/2\pi=350\pm3$ MHz ($g_{\rm{JC}}/2\pi=351\pm3$ MHz). See also Fig.~\ref{figS4}(a).
		(c) Linecut representing $|S_{11}|(\omega_{\rm{p}}/2 \pi)$ taken along the black arrows in (b). 
		The orange line represents a fit to a JC master equation model.
%		from which we extract $g/2\pi \sim 373.4 \pm 0.6 ~\rm{MHz}$ [$g/2\pi \sim 628.8 \pm 1.4 ~\rm{MHz}$] and $\Gamma_{2}/2\pi \sim 56.3 \pm0.2\, \rm{MHz}$ [$\Gamma_{2}/2\pi \sim 149 \pm\, 2 \rm{MHz}$] in correspondence of $\eta \sim 0.50 \pm 0.14$ [$\eta \sim 0.72 \pm 0.08$] 
		The resonator losses are $\kappa_{int}/2\pi=19.5 \,\rm{MHz}$ and $\kappa_{ext}/2\pi=4.3\pm0.1\,\rm{MHz}$.
% from which we extract 
%		$\tunneling_{\rm{R}}/2\pi = 5.581 \pm 0.008$ GHz ($\tunneling_{\rm{JC}}/2\pi= 5.612 \pm 0.007$ GHz),
%		$\omega_{r,\rm{R}}/2\pi \sim 5.6612 \pm 0.0003~\rm{GHz}$ ($\omega_{r,\rm{JC}}/2\pi \sim 5.6635 \pm 0.0002~\rm{GHz}$)  
%		and $g_{\rm{R}}/2\pi=619\pm4$ MHz ($g_{\rm{JC}}/2\pi=629\pm4$ MHz). See also Fig.~\ref{figS4}(b).
	}
	\label{fig6:Ultrastorngcoupling_top}
\end{figure*}

Figure \ref{fig6:Ultrastorngcoupling_top}(a) show a study of a DQD configuration, distinct from what displayed in Fig.~\ref{fig6:Ultrastorngcoupling} inthe main text. This new configuration characterized by $\eta \sim 0.5$ has been obtained by in-situ tuning the DQD dipole strength as described in the main text..
The red (blue) line in Fig.~\ref{fig6:Ultrastorngcoupling_top}(b) represents a fit to the data obtained using the Rabi (JC) model from which we can extract $g_{\rm{R}}/2\pi = 350 \pm 3 ~\rm{MHz}$ ($g_{\rm{JC}}/2\pi = 351 \pm 2~\rm{MHz}$).
A fit to a master equation model [solid orange line in Fig.~\ref{fig6:Ultrastorngcoupling_top}(c)] to the Rabi mode spectrum, obtained by changing the probe frequency along the DQD detuning value indicated by the black arrows in Fig.~\ref{fig6:Ultrastorngcoupling_top}(b),
yields a splitting of $\qubitResonatorCoupling/2\pi \sim 373.4 \pm 0.3~ \rm{MHz}$, with a DQD charge decoherence of $\Gamma_{2}/2\pi \sim 56.3 \pm 0.2~ \rm{MHz}$. 
%($\Gamma_{2}\sim 44.4 \pm2.0 ~\rm{MHz}$).
For this DQD electrostatic configuration the system is in the strong coupling regime ($g > \kappa/2 + \Gamma_2$) but presents a $g_{\rm{R,JC}}/\omega_{\rm{r}} \sim 0.062$, which despite being very high for a DQD-resonator hybrid device does not promote the system in the USC regime.

\begin{figure*}[!htb]
\includegraphics[width=\textwidth]{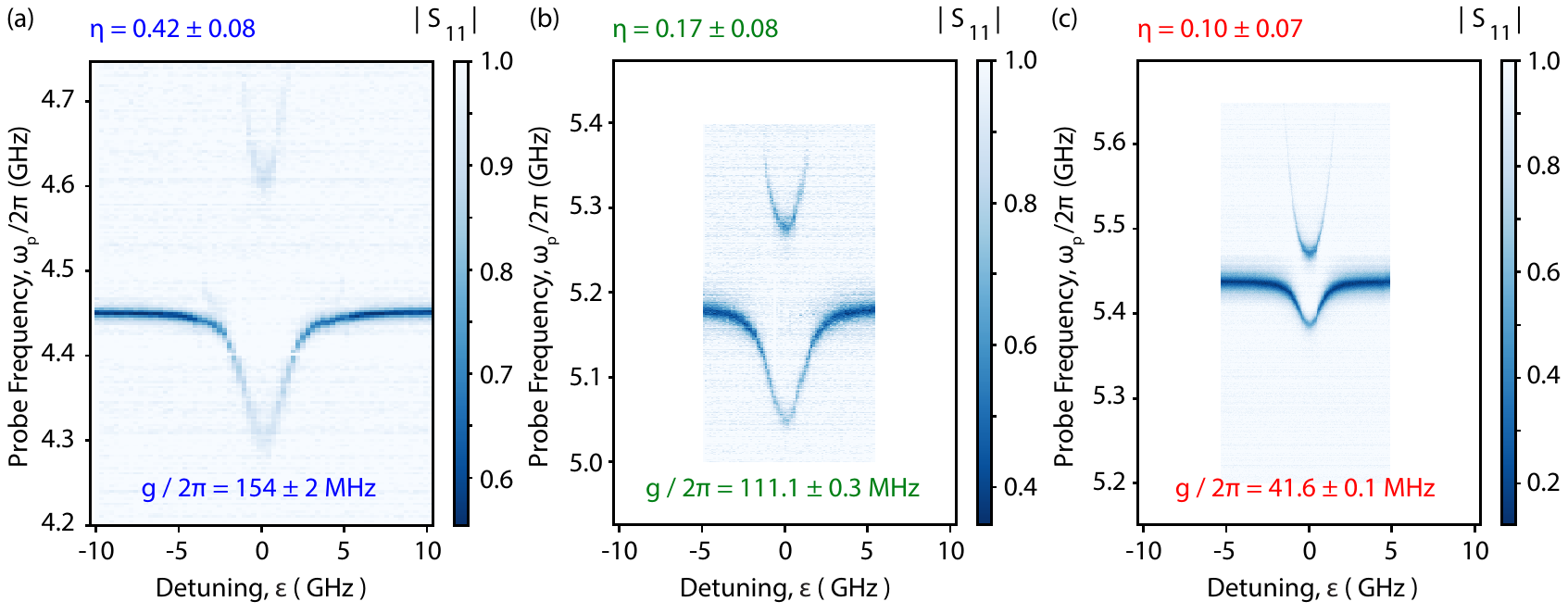}
\caption{
Response of the reflectance amplitude $|S_{11}|$ of the SQUID array resonator \emph{vs.} DQD detuning $\detuning$ in correspondence of three distinct dipole strengths (a) $\eta \sim 0.42 \pm 0.08$ (blue), (b) $\eta \sim 0.17 \pm 0.08$ (green) and (c) $\eta \sim 0.10 \pm 0.07$ (red) [the corresponding DQD
charge stability diagrams are reported in Fig.~\ref{fig2:four_stability_diagramss}(e), (d) and (c)].
%by monitoring the change in resonator phase difference $\Delta\phi$. 
The three resonant spectrums are obtained by tuning the SQUID array in resonance with the DQD excitation frequency for $\detuning=0$.
% $(\nu_\mathrm{r} = \nu_\mathrm{q})$. 
Data already reported in Fig.~\ref{fig3:Display_porperties}(a-c).
}
\label{figS3}
\end{figure*}
\begin{figure*}[!htb]
	\includegraphics[width=\textwidth]{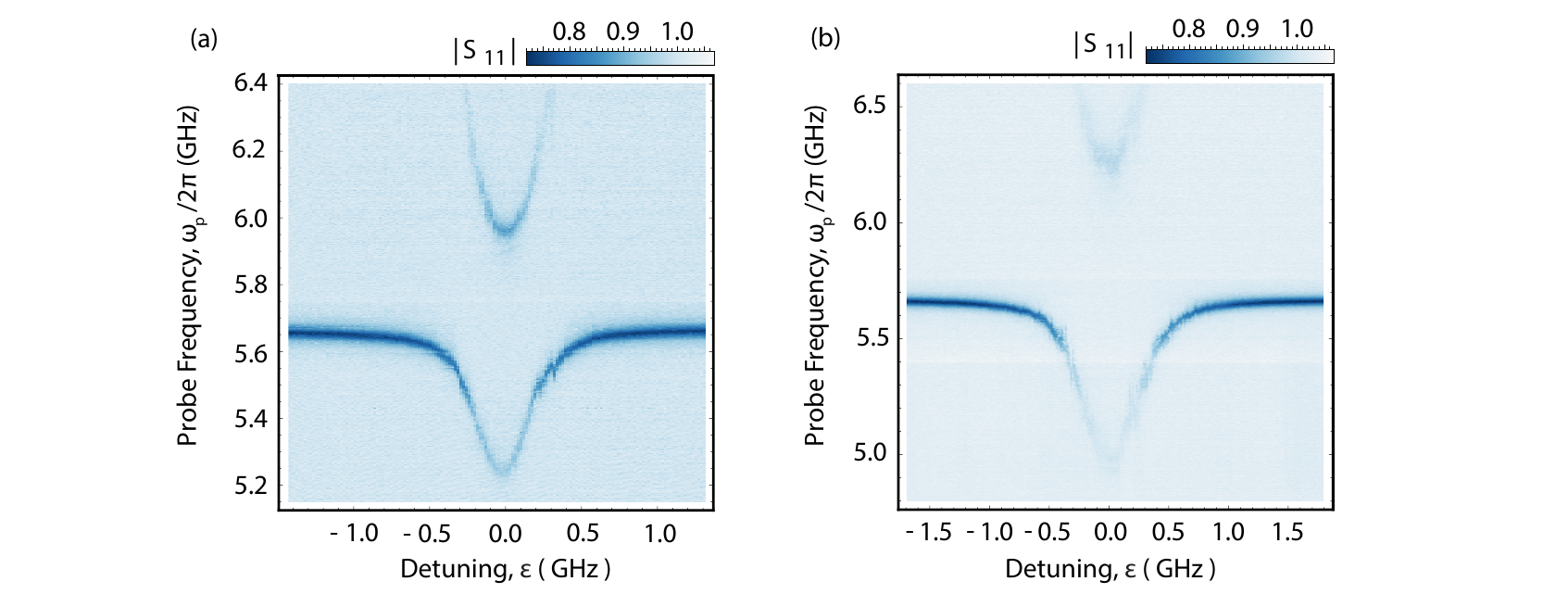}
	\caption{
	Response of the reflectance amplitude $|S_{11}|$ of the Josephson junction array resonator \emph{vs.} DQD detuning $\detuning$ in correspondence of a DQD configuration characterized by parameter (a) $\eta \sim 0.50 \pm 0.14$ and (b) $\eta \sim 0.72 \pm 0.08$.
	Data already reported in Fig.~\ref{fig6:Ultrastorngcoupling_top}(b) and Fig.~\ref{fig6:Ultrastorngcoupling}(b).
	}
	\label{figS4}
\end{figure*}

%\end{widetext}
\end{document}